\documentclass{aa}  

\usepackage{graphicx}
\usepackage{epstopdf}
\usepackage{color}
\usepackage{amsmath}
\usepackage{subcaption}

\usepackage{txfonts}
\newcommand{\integral}{{\it INTEGRAL}}
\newcommand{\bepposax}{{\it BeppoSAX}}

\newcommand{\swift}{{\it Swift}}
\newcommand{\nustar}{{\it NuSTAR}}
\newcommand{\swiftbat}{{\it Swift}/BAT}
\newcommand{\swiftxrt}{{\it Swift}/XRT}
\newcommand{\maxi}{{\it MAXI}}
\newcommand{\agile}{{\it AGILE}}
\newcommand{\fermi}{{\it Fermi}}
\newcommand{\fermilat}{{\it Fermi}-LAT}

\begin{document}

   \title{The hypersoft state of Cygnus X--3}

   \subtitle{A key to jet quenching in X-ray binaries?}

   \author{K.~I.~I.~Koljonen
          \inst{1,2}
          \and
          T.~Maccarone\inst{3} \and
          M.~L.~McCollough\inst{4} \and
          M.~Gurwell\inst{4} \and
          S.~A.~Trushkin\inst{5,6} \and
          G.~G.~Pooley\inst{7} \and
          G.~Piano\inst{8} \and
          M.~Tavani\inst{8,9,10}
          }

   \institute{Finnish Centre for Astronomy with ESO (FINCA), University of Turku, V\"ais\"al\"antie 20, 21500 Piikki\"o, Finland \\
              \email{karri.koljonen@utu.fi}
         \and
             Aalto University Mets\"ahovi Radio Observatory, PO Box 13000, FI-00076 Aalto, Finland \\
         \and    
             Department of Physics and Astronomy, Texas Tech University, Box 41051, Lubbock, TX 79409-1051, USA \\
         \and
             Harvard-Smithsonian Center for Astrophysics, Cambridge, MA 02138 US \\
         \and
             Special Astrophysical  Observatory RAS, Nizhnij Arkhyz, 369167, Russia \\
         \and
             Kazan Federal University, Kazan, 420008, Russia \\
         \and
             University of Cambridge, Astrophysics, Cavendish Laboratory, Cambridge CB3 0HE, UK \\      
         \and
             INAF-IAPS, Via del Fosso del Cavaliere 100, I-00133, Roma, Italy \\
         \and
             INFN Roma Tor Vergata, Via della Ricerca Scientifica 1, I-00133, Roma, Italy \\
         \and
             Dipartimento di Fisica, Universit\`a di Roma ``Tor Vergata'', Via Orazio Raimondo 18, I-00173, Roma, Italy \\               
             }

   \date{Received ; accepted }

 
  \abstract
   {Cygnus X-3 is a unique microquasar in the Galaxy hosting a Wolf-Rayet companion orbiting a compact object that most likely is a low-mass black hole. The unique source properties are likely due to the interaction of the compact object with the heavy stellar wind of the companion.}
   {In this paper, we concentrate on a very specific period of time prior to the massive outbursts observed from the source. During this period, Cygnus X-3 is in a so-called hypersoft state, where the radio and hard X-ray fluxes are found to be at their lowest values (or non-detected), the soft X-ray flux is at its highest values, and sporadic gamma-ray emission is observed. We will utilize multiwavelength observations in order to study the nature of the hypersoft state.}
   {We observed Cygnus X-3 during the hypersoft state with \swift\/ and \nustar\/ in the X-rays and SMA, AMI-LA, and RATAN-600 in the radio. We also considered X-ray monitoring data from \maxi\/ and $\gamma$-ray monitoring data from \agile\/ and \fermi.}
   {We found that the spectra and timing properties of the multiwavelength observations can be explained by a scenario where the jet production is turned off or highly diminished in the hypersoft state and the missing jet pressure allows the wind to refill the region close to the black hole. The results provide proof of actual jet quenching in soft states of X-ray binaries.}
   {} 

   \keywords{Accretion, accretion disks -- binaries: close -- stars: individual: Cygnus X--3 -- stars: winds, outflows -- X-rays: binaries}

   \maketitle
%

\section{Introduction} \label{introduction}

Cygnus X--3 (Cyg X--3) is a unique X-ray binary (XRB) in the Galaxy hosting a Wolf-Rayet (WR) companion orbiting a compact object (\citealt{vankerkwijk92}; Koljonen et al. 2017). It is a persistent, bright source in the radio as well as in the X-rays, featuring radio flux density levels around 100 mJy \citep[e.g.][]{waltman94} and $2-8 \times 10^{-8}$ erg cm$^{-2}$ s$^{-1}$ bolometric X-ray flux \citep[e.g.][]{hjalmarsdotter08} $\sim$50\% of the time. For a distance estimate of 7.4 kpc \citep{mccollough16} these correspond to luminosities 10$^{31}$-10$^{32}$ erg s$^{-1}$ and 1--5 $\times$ 10$^{38}$ erg s$^{-1}$, respectively (for the second best distance solution of 10.2 kpc the luminosities are a factor of two higher; see \citealt{mccollough16} for details on the distance estimation). Occasionally, Cyg X--3 undergoes giant radio outbursts, during which the radio flux density levels can reach 20 Jy \citep{waltman96} making Cyg X--3 the brightest Galactic radio source. During the outbursts, a one-sided relativistic jet with multiple knots has been resolved with Very Long Baseline Interferometry (VLBI; \citealt{mioduszewski01,millerjones04,tudose07}). The jet morphology implies that the jet axis lies close to our line-of-sight ($<$14 degrees; \citealt{mioduszewski01,millerjones09}). Contrary to other XRBs, these outbursts are seen when the source makes transitions from the high/soft state to the low/hard state \citep{szostek08b,koljonen10}. The outbursts are preceded by a radio quiet period (down to $\sim$ 1 mJy; \citealt{waltman96, fender97, koljonen10}), where the X-ray spectra are found to be at their softest (so-called ``hypersoft'' state; \citealt{koljonen10}). $\gamma$-ray emission is most often detected from Cyg X-3 when the source is transiting to/from the hypersoft state, and in occasion during the hypersoft state when connected to minor radio flaring episodes \citep{tavani09,fermi09,corbel12,bulgarelli12}. It has been suggested that the unique source properties, namely the peculiar X-ray spectra and $\gamma$-ray emission, are due to the short orbital separation (2--3$\times 10^{11}$ cm) coupled with the interaction of the compact object with the high-density stellar wind and photon field ($L_{WR} \sim 4-6 \times 10^{38}$ erg s$^{-1}$) of the WR companion where the compact object is embedded \citep{paerels00,szostek08a,zdziarski10,dubus10,piano12,zdziarski12}.

Previously, a similar X-ray spectrum which was defined as hypersoft state spectrum in \citet{koljonen10} has been observed from Cyg X-3 in \citet{smale93} using the Broad-Band X-ray Telescope, in \citet{beckmann07} using \integral\/, and in \citet{szostek05} using \bepposax. This state is characterized by the lack of iron lines (which we will show here, is an orbital effect), and pure blackbody emission with a temperature of 1.1 keV (which alternatively can be fit by a Comptonized accretion disk spectrum where the scattering electron temperature is close to the seed photon temperature; \citealt{koljonen10}) and a faint power law tail (with a power law index $\Gamma \, \sim$2) spanning from 20 keV onwards. During the hypersoft state, the \swiftbat\/ hard X-ray flux (15--50 keV band) is consistent with zero. At the same time, the \maxi\/ lightcurve is at its maximum: 1--2 cts/cm$^{2}$/s in the 2--20 keV band. 

In addition to Cyg X-3, a similar hypersoft state X-ray spectrum has been reported from GRO J1655--40 \citep{uttley15} and Swift J1753.5--0124 (\citealt{shaw16}; albeit here with much lower blackbody/disk temperature). GRO J1655--40 is expected to have a strong magnetically driven accretion disk wind and high inclination \citep{orosz97,miller08,neilsen12}. It has been suggested that the unusually soft spectrum is a consequence of Compton thick and ionized disk wind obscuring the X-ray source in the line-of-sight \citep{uttley15,shidatsu16,neilsen16}. Swift J1753.5--0124 also presented a similar, unusual soft state spectrum, where the accretion disk emission is presumably Compton scattered by a disk atmosphere \citep{shaw16}. The observation coincided with a prolonged state of a quenched radio emission ($<$ 21 $\mu$Jy) and zero \swiftbat\/ flux \citep{rushton16}. The hypersoft state also bears similarities with ultraluminous supersoft sources that have very soft, thermal, or steep power-law ($\Gamma\sim3-4$) spectra, which produce disk winds due to super-Eddington accretion \citep{earnshaw17}. 

In this paper, we study the hypersoft state of Cyg X-3 preceding the 2016 and 2017 outburst episodes with \nustar\/, \swift\/ and \maxi\/ in the X-rays, and with RATAN-600/SMA/AMI-LA in the radio together with supporting monitoring observations in the $\gamma$-rays with \agile\/ and \fermi.. In Section \ref{observations}, we introduce the radio/X-ray/$\gamma$-ray observations and outline their data reduction processes. In Section \ref{results}, we present the multiwavelength view of the 2016/2017 outbursts and concentrate on the radio/X-ray/$\gamma$-ray properties (spectra and variability) during the hypersoft state. In Section \ref{discussion}, based on our observations, we discuss a possible scenario to explain the radio/X-ray/$\gamma$-ray properties by assuming that during the hypersoft state the jet turns off. In Section \ref{conclusions}, we conclude our findings.        

\section{Observations and data reduction} \label{observations}

We have radio monitoring programs on Cyg X-3 running daily/weekly observations with RATAN-600 and AMI-LA, with higher cadence observing during the outburst episodes. These together with the X-ray monitoring (see below) were used to alert us about the start of the hypersoft state. After triggering, we used ToO/DDT proposals with the SMA, \swift\/, and \nustar\/ to study the radio and X-ray properties in more detail. Below, we briefly outline the observations and data reduction processes for each observatory/instrument.

\subsection{Radio data}

\subsubsection{SMA}

The Submillimeter Array (SMA), located just below the summit of Maunakea, Hawaii, is an interferometer consisting of eight 6-m diameter antennas configurable to cover baselines ranging from 8 m to 509 m, with receivers capable of covering frequencies from $\sim$190 to 420 GHz. Observations in 2016 were obtained with a hybrid setup of correlators, covering 8 GHz of bandwidth in each of two sidebands from a single polarization receiver (16 GHz total continuum bandwidth), with a mean frequency of 225.5 GHz. In 2017, observations were made with the new SWARM correlator covering 8 GHz of bandwidth in each of two sidebands from two orthogonally polarized receiving systems (32 GHz total continuum bandwidth), with a mean frequency of 220--230 GHz. In both years, time spent on source varied between several minutes to several hours. The flux density scale was referenced to the nearby source MWC349A for all observations, providing an absolute flux uncertainty of roughly 5\%; however, the relative uncertainty from day to day is much smaller, below 1\%.

\subsubsection{AMI-LA}

The AMI Large Array, near Cambridge (UK), is the rebuilt and re-engineered 5-km telescope array \citep{zwart08}. It has eight 13-m antennas, and operates in the 13 to18 GHz band, with a high-resolution digital correlator. We collected data using the full bandwidth with a centre frequency of 15.5 GHz and typically observing the source from $\sim$10-minute up to few hours at a time. We use an interleaved calibrator to keep the phases as nearly correct as we can and calibrate the flux density measurements using a calibrator source with a typical variability of 5\% in amplitude.   

\subsubsection{RATAN-600}

RATAN-600 is a radio telescope of the Special Astrophysical Observatory of the Russian Academy of Science located in Nizhnij Arkhyz, Russia. The antenna consists of a 576-m circle of 895 reflecting elements which can be used as four independent sectors. Several different feed cabins with secondary mirrors can collect the reflected emission for the primary receivers. The receiver complex consists of several radiometers in the wavelength range spanning from 0.6 GHz to 30 GHz. We used data that were taken with the northern sector at 4.6 GHz, and with the southern sector at 4.8 GHz. The errors on the flux density measurements are $\sim$ 5 mJy and $\sim$ 10 mJy for a level of 100 mJy or 3\% and 5\% at fluxes $>$ 1 Jy for the northern and the southern sectors, respectively. In addition, at selected times we took data in other bands: 2.3, 7.7, 11.2 and 21.7 GHz as well \citep[see data in][]{trushkin17b,trushkin17}.

\subsection{X-rays}

We used X-ray monitoring data from \maxi/GSC \citep{matsuoka09} and \swiftbat\/ \citep{krimm13} and obtained the daily and orbital 2--10 keV and 15--50 keV fluxes from their web interfaces. In addition to the monitoring observations, five pointings were observed with \swift\/ during the outbursts when the source was in the hypersoft state. In addition, one pointing with \nustar\/ was observed during the decay of a minor flare observed in the middle of the hypersoft state before the major flare ejection in 2017. 

\subsubsection{\swift}

The \swiftxrt\/ windowed timing (WT) mode data were processed using {\sc xrtpipeline}. Sources that are heavily absorbed (including Cyg X-3) show residuals in the WT spectra at low energies (0.4--1.0 keV), when using grades greater than 0. Thus, we extract only grade 0 spectra. In addition, multi-pixel events sometimes cross the 10-row binning boundaries of the WT mode and become split during the readout process, which affects the redistribution tail seen in this mode. In order to correct this effect, we use the position-dependent WT redistribution matrices, that correct the spectra at low energies. With these extraction steps, we are able to include the 0.4--1.0 keV data for the X-ray spectra model fitting. The source and background spectra together with the ancillary response are extracted with {\sc xrtproducts}. Depending on the source count rate, we extract the source data from an annulus and background from a circle. For the X-ray modeling we bin the data to S/N$=$10 in the band 0.5--1.5 keV, S/N$=$20 in the band 1.5--1.8 keV, S/N$=$30 in the band 1.8--3.0 keV, S/N$=$40 in the band 3--6 keV and S/N$=$30 in the band 6--10 keV. This gives a roughly equal number of energy bands throughout the spectrum. 

\subsubsection{\nustar} \label{nustar_red}

We reduced the data from both detectors (FMPA/FMPB) using \textsc{nupipeline}. We used a circular source region with a 60-arcsec radius centered on the location of Cyg X-3, and a circular background region with a 90-arcsec radius that was selected from a sourceless region in the detector image. The pipeline was run with parameters SAAMODE$=$`optimized' and TENTACLE$=$`yes'. The former requires the presence of an increase in the CdZnTe detector event count rates simultaneous to the observed shield single rates increase, and the latter allowing identification and flagging of time intervals in which the CdZnTe detector events count rates show an increase when the spacecraft is entering the South Atlantic Anomaly (SAA). We extracted a spectrum from both detectors using the whole pointing ($\sim$20 ksec), and in addition, divided the pointing to 18 spectra with exposures of $\sim$1 ksec in order to study the spectral evolution (e.g. orbital modulation). We also extracted lightcurves from soft (3--10 keV) and hard (20--60 keV) energy bands with 50-second and 100-second time bins, respectively. In addition, we extracted 0.5 sec lightcurve from the full 3--79 keV band to study the fast variability. 

For X-ray modeling we bin the data to S/N$=$20 in the band 3--10 keV, S/N$=$10 in the band 10--20 keV and S/N$=$5 in the band 20--79 keV. This gives a roughly equal number of energy bands throughout the spectrum. 

For the timing analysis, we calculated the cospectra from the continuous data segments of the 0.5 sec lightcurves of both detectors. The cospectrum is the real part of the complex cross spectrum that is the Fourier transform of a time series (in this case the lightcurve from FMPA) multiplied with the complex conjugate of the Fourier transform of another time series (the lightcurve from FMPB). The cospectrum is used to mitigate instrumental effects, especially dead time that is present in the \nustar\/ lightcurves \citep{bachetti15a}, and it can be used as a proxy for the (white-noise-subtracted) power density spectrum (PDS). We calculate the cospectra using MaLTPyNT \citep{bachetti15b}, which are then used to make an average cospectrum.

\begin{table}
\caption{X-ray observations} \label{obs}
\begin{tabular}{lcccc}
\centering
\textbf{Swift} & Start Date & Exp. & Start Phase \\
Pointing & MJD & [ksec] \\
\hline
00031922032 & 57626.33859 & 0.98 & 0.52 \\
00031922033 & 57629.65261 & 0.97 & 0.11 \\
00059162001 & 57801.61197 & 1.55 & 0.23 \\
00059163001 & 57801.67723 & 1.55 & 0.55 \\
00059164001 & 57801.74321 & 1.55 & 0.88 \\
\hline
\textbf{NuSTAR} & Start Date & Exp. & Elapsed \\
Pointing & MJD & [ksec] & [ksec] \\
\hline
90202051002 & 57814.23476 & 14.3 & 43.7 \\
\end{tabular}
\end{table}

\subsection{$\gamma$-rays}

We used $\gamma$-ray monitoring data from the GRID instrument \citep{barbiellini02,prest03} onboard \agile\/ \citep{tavani09b}. The data were collected during both outburst episodes including the hypersoft states. We registered only those events which present TS values over 6. The data reduction and analysis methods are described in \citet{piano12}. 

In addition to \agile\/ data, we used reported \fermilat\/ detections \citep{cheung16} for the 2016 outburst episode, and the data from \fermilat\/ monitored source list\footnote{\url{https://fermi.gsfc.nasa.gov/ssc/data/access/lat/msl_lc/}} for the 2017 outburst event. 

\section{Results} \label{results}

\subsection{Multiwavelength overview}

\begin{figure*}
\centering
\begin{subfigure}{.49\textwidth}
  \centering
  \includegraphics[width=\linewidth]{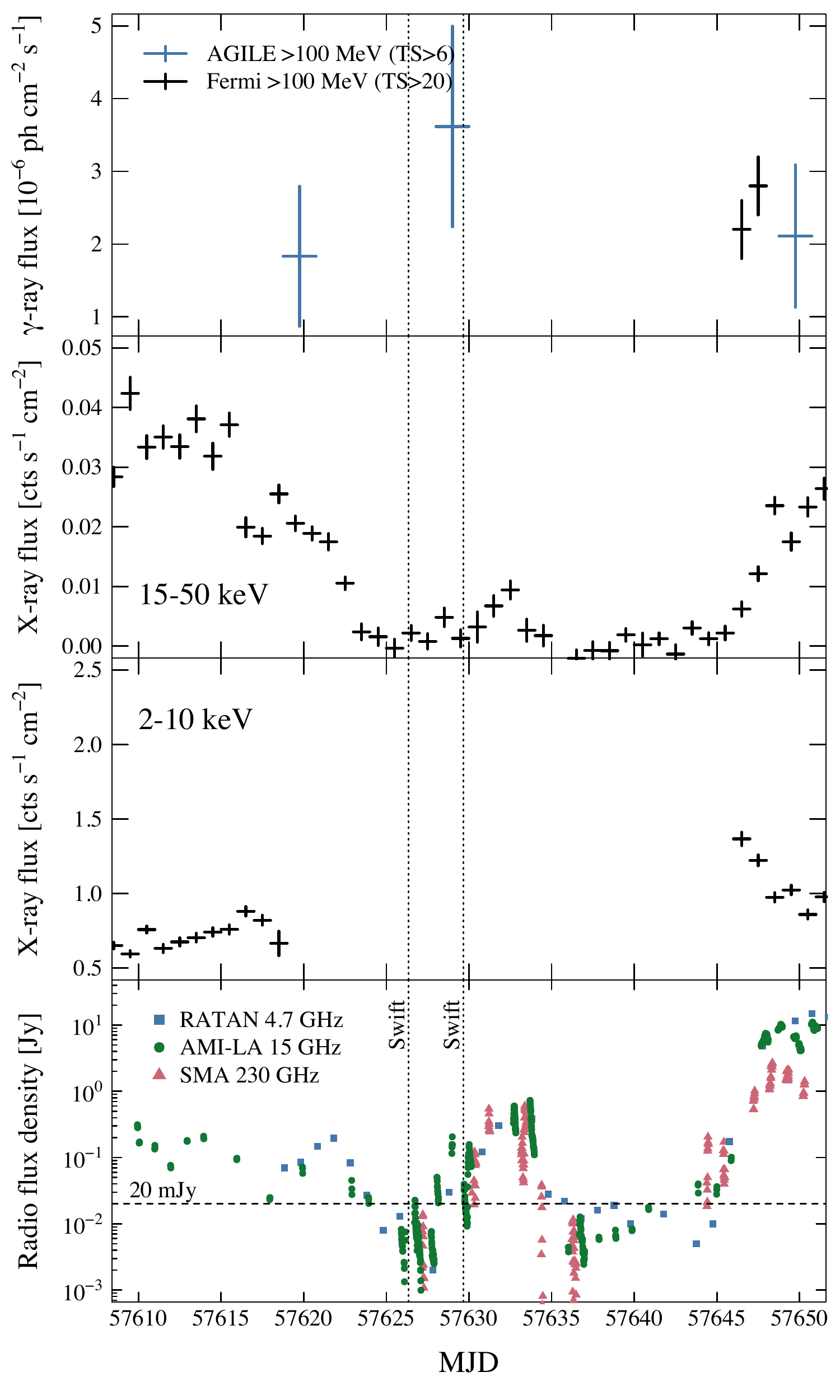}
\end{subfigure}
\begin{subfigure}{.49\textwidth}
  \centering
  \includegraphics[width=\linewidth]{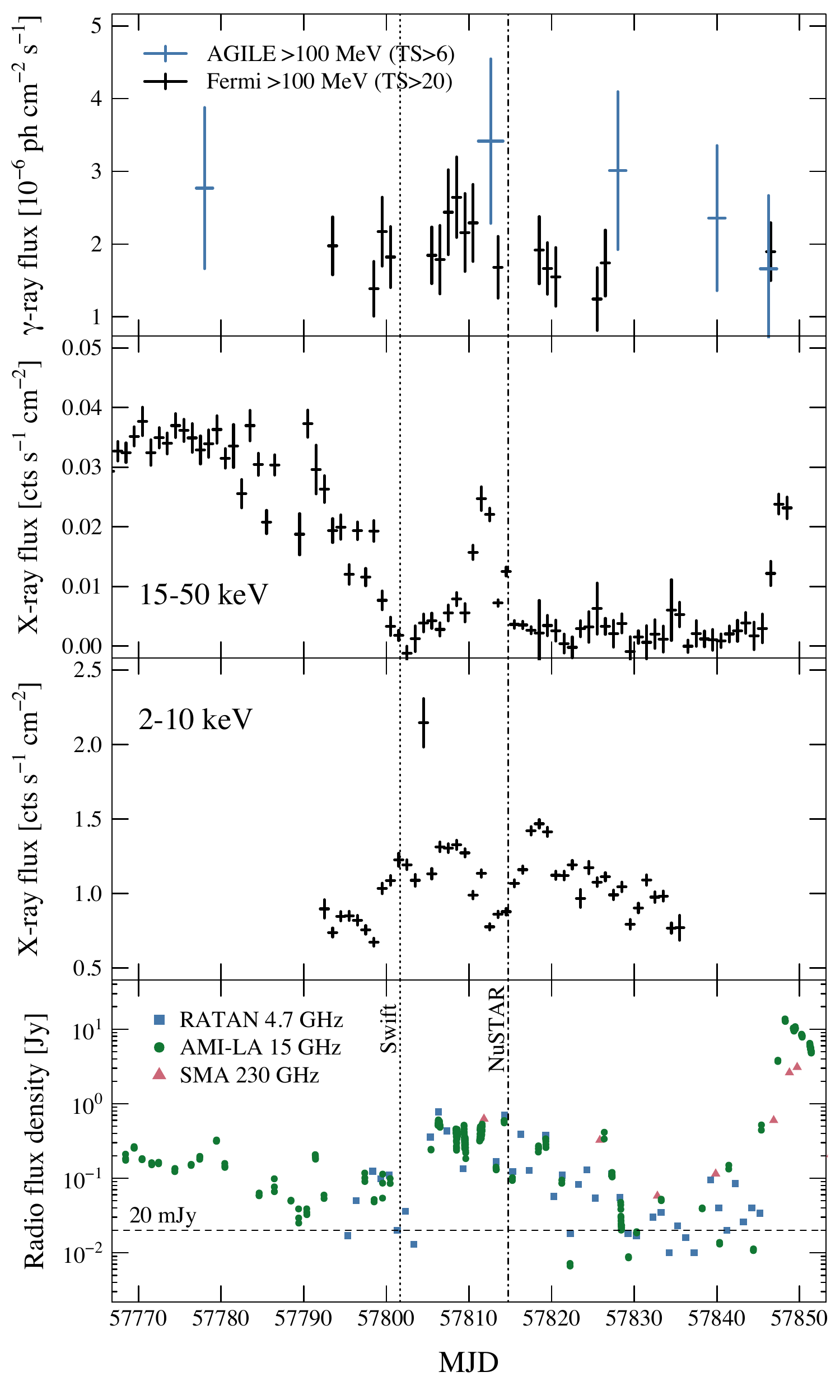}
\end{subfigure}
\caption{\textit{Left:} $\gamma$-ray (\agile\/ and \fermi), X-ray (\swiftbat\/ and \maxi) and radio (RATAN, AMI-LA, and SMA) monitoring of the 2016 outburst episode. \textit{Right:} The same monitoring campaign for the 2017 outburst. Hypersoft state occurs when the \swiftbat\/ 15--50 keV count rate is approximately 0 cts s$^{-1}$ cm$^{-2}$, and radio flux is below 20 mJy. Usually, a prolonged ($\sim$10--30 days) hypersoft state occurs before outbursts, that can exhibit also brief (few days) flares in the radio and hard X-rays, as happened in both 2016 and 2017 outbursts. $\gamma$-ray emission is usually attributed to changes to/from the hypersoft state (during the onset, and/or during the hard X-ray/radio flare, and/or during the onset of the outburst). Dotted and dot-dashed lines denote \swift\/ and \nustar\/ pointings, respectively.}
\label{overview}
\end{figure*}

During the multiwavelength campaign of Cyg X-3 in 2016 and 2017, the source presented similar behavior for both outburst episodes (Fig. \ref{overview}). The hypersoft state preceded the major outburst for a length of $\sim$20 days and $\sim$45 days for 2016 and 2017, respectively, interspersed by moderate radio (radio flux densities reaching 1 Jy) and hard X-ray flaring (correlated with the radio). On average, the 2017 outburst presented slightly higher radio flux densities throughout the outburst episode including the hypersoft state, where the (frequency-independent) radio flux density varied between 10--100 mJy during zero flux from \swiftbat\/, while in 2016 the radio flux level was lower at 1--20 mJy.  

Several ATels reported $\gamma$-ray detections during the 2016/2017 outburst episodes. In 2016, a $\gamma$-ray flare was observed at the onset of the minor radio flare \citep{piano16,cheung16}. In 2017, three flares were reported: one in transit into the hypersoft state, one during a minor radio flare, and one at the onset of the major radio flare \citep{loh17a,loh17b,piano17a,piano17b}.  

\subsection{Radio properties}

\begin{figure*}
 \centering
 \includegraphics[width=\linewidth]{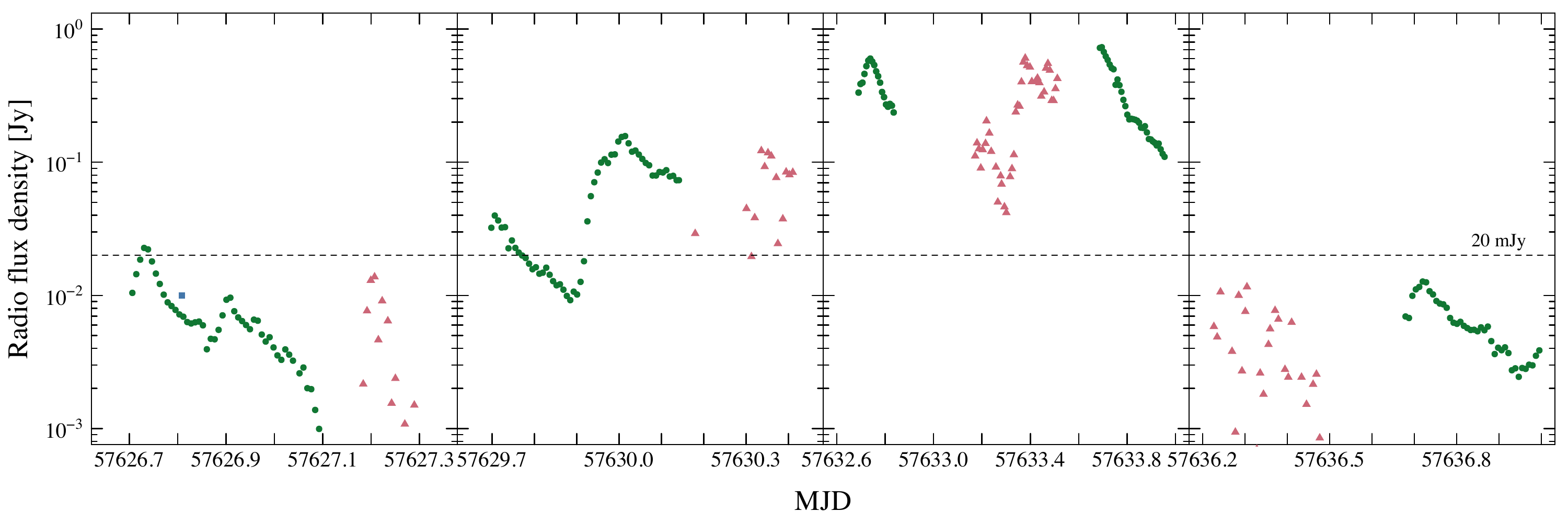}
 \caption{Intra-day radio lightcurves (green: 15 GHz, red: 230 GHz) from the 2016 hypersoft state (first and last panel) and minor radio flare (middle panels). The colors are same as in Fig. \ref{overview}. The 230 GHz lightcurve seems to present higher variability (see Fig. \ref{sma_detail} and Fig. \ref{radio_fvar}). The 15 GHz lightcurve present some well defined flares that last $\sim$0.1 days, indicating a size scale of 3$\times10^{14}$ cm (well beyond the orbital scale).}
 \label{radio_detail}
\end{figure*}

\begin{figure}
 \centering
 \includegraphics[width=\linewidth]{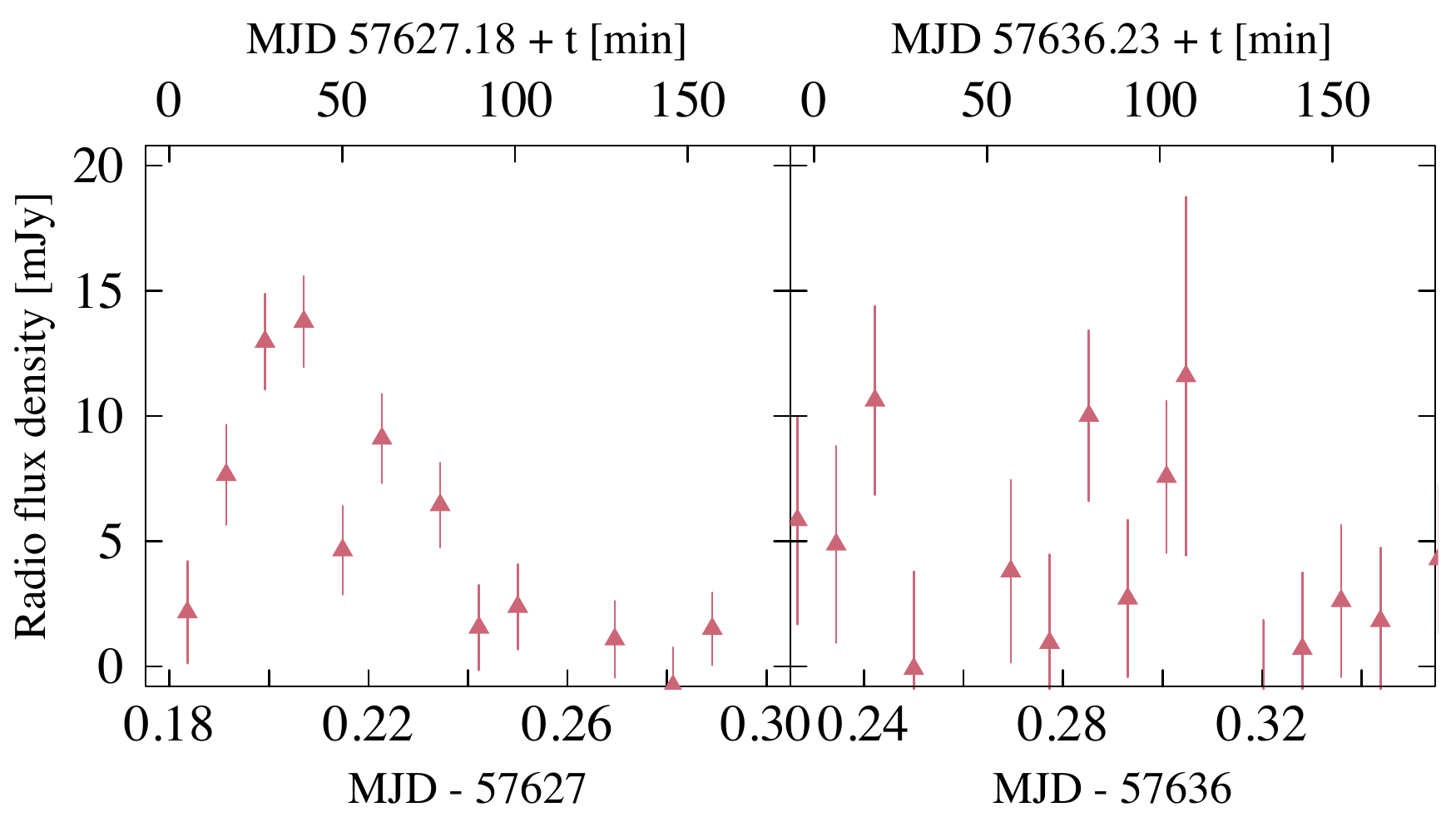}
 \caption{The 230 GHz lightcurve during hypersoft state presenting rapid flickering with $\sim$20--50-minute timescale, indicating a size scale of 3--9$\times10^{13}$ cm (well beyond the orbital scale).}
 \label{sma_detail}
\end{figure}
 
Intra-day radio lightcurves during the hypersoft state show strong radio modulation with $\sim$0.1 day flares in 15 GHz and fast ($\sim$20--50 minutes) flickering in 230 GHz (Figs. \ref{radio_detail}, \ref{sma_detail}). These variability time scales correspond to emission regions of 3$\times10^{14}$ cm and 3--9$\times10^{13}$ cm, respectively, and both lie well beyond the orbital separation that for 4.8-hour orbital period and 10--20 M$_{\odot}$ total mass of the binary is 2--3$\times 10^{11}$ cm. 

Assuming that the jet radio emission is attenuated by free-free absorption in the WR stellar wind \citep{fender95,waltman96}, the radio photosphere $R$ for a given frequency $\nu$ can be approximated as \citep{wright75,waltman96}: 

\begin{equation}
R_{\tau\sim1} = 1.75 \times 10^{28} \gamma^{1/3} g_{ff}^{1/3} Z^{2/3} T^{-1/2} \Bigg( \frac{\dot{M}}{\mu v_{\infty} \nu} \Bigg )^{2/3} \mathrm{cm},
 \label{photosphere}
\end{equation} 

\noindent where the free-free Gaunt factor, $g_{ff}$, is derived from the relation \citep{leitherer91}: 

\begin{equation}
g_{ff} = 9.77 \times \Bigg[ 1.0 + 0.13 \, \mathrm{log} \Bigg( \frac{T^{3/2}}{\nu Z} \Bigg) \Bigg].
\end{equation}

Using the following parameters for the stellar wind (Koljonen et al. 2017): the mass-loss rate of $\dot{M} = 10^{-5} \, M_{\odot}$/yr, the temperature of $T = 45000$ K, the terminal velocity of the stellar wind of $v_{\infty}=1000$ km/s, the rms ionic charge of $Z=1$, the mean number of free electrons per nucleon of $\gamma=1$, and the mean atomic weight per nucleon of $\mu=3.6$, we get 5$\times10^{13}$ cm and 7$\times10^{12}$ cm for the 15 GHz and 230 GHz photosphere radii. Thus, these values are consistent with the variability timescales.

\begin{figure}
 \centering
 \includegraphics[width=\linewidth]{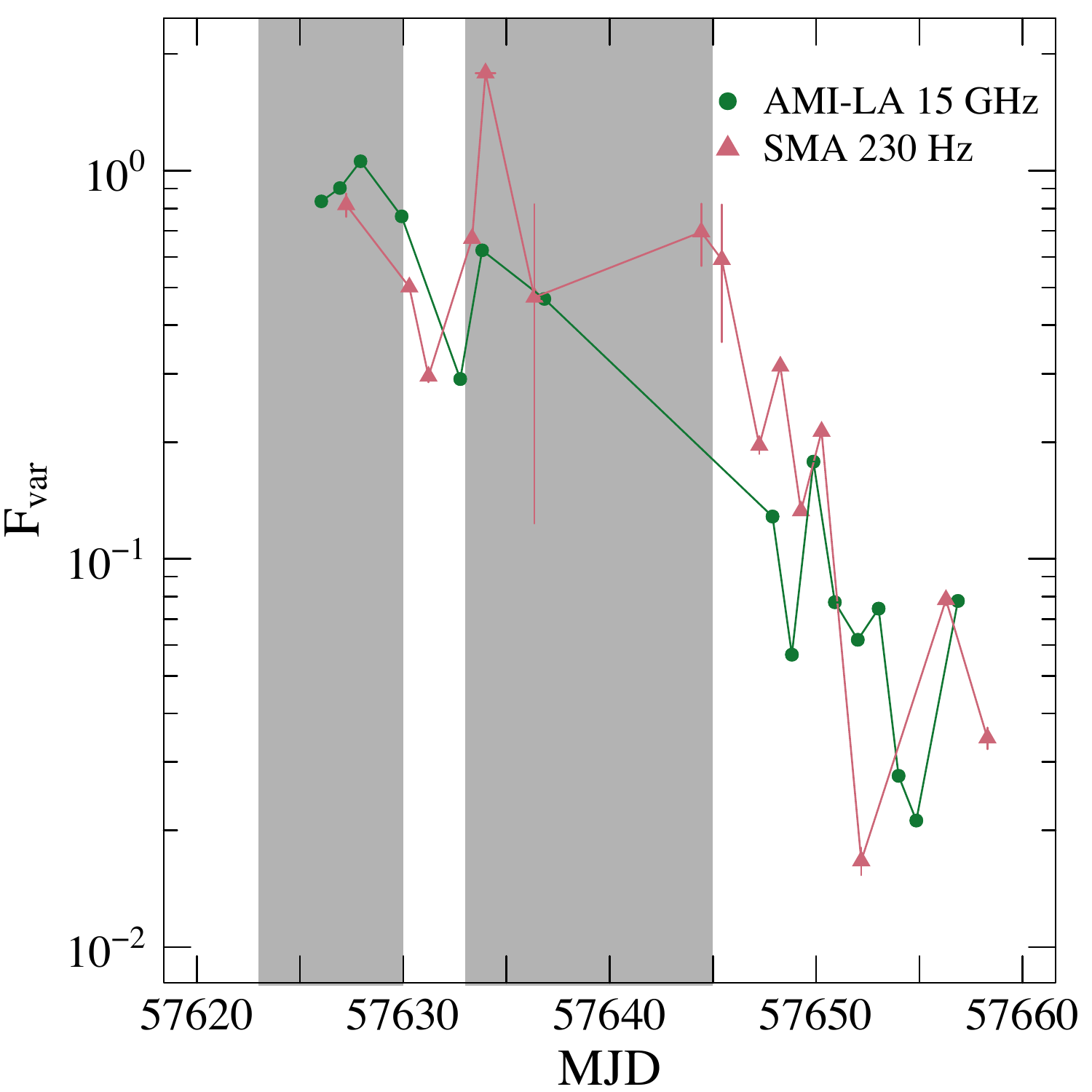}
 \caption{The fractional variability amplitude (coefficient of variation) of the radio lightcurves (green: 15 GHz, red: 230 GHz). The evolution in the 15 GHz as well as in 230 GHz is similar. Also, during the hypersoft state (around MJD 57625 and MJD 57635, colored areas) the fractional variability amplitude is at maximum and presenting variations exceeding the mean, while during the flares it is markedly lower.}
 \label{radio_fvar}
\end{figure}

We calculated the fractional variability amplitude (\citealt{vaughan03}, and references therein; also known as the coefficient of variation) from the intraday 15 GHz and 230 GHz radio fluxes (Fig. \ref{radio_fvar}; we disregarded the observations that presented less than 10 subsequent radio measurements). The amplitude of the radio variability is at its maximum during the hypersoft state, and the radio flux exhibits variations that exceed the average flux, while during radio flares it decreases -- more strongly for the outburst than during moderate radio flaring. If the radio variability is connected to the light-crossing time of the emission region, this indicates that the hypersoft state presents more compact emission regions. 

\begin{figure}
 \centering
 \includegraphics[width=\linewidth]{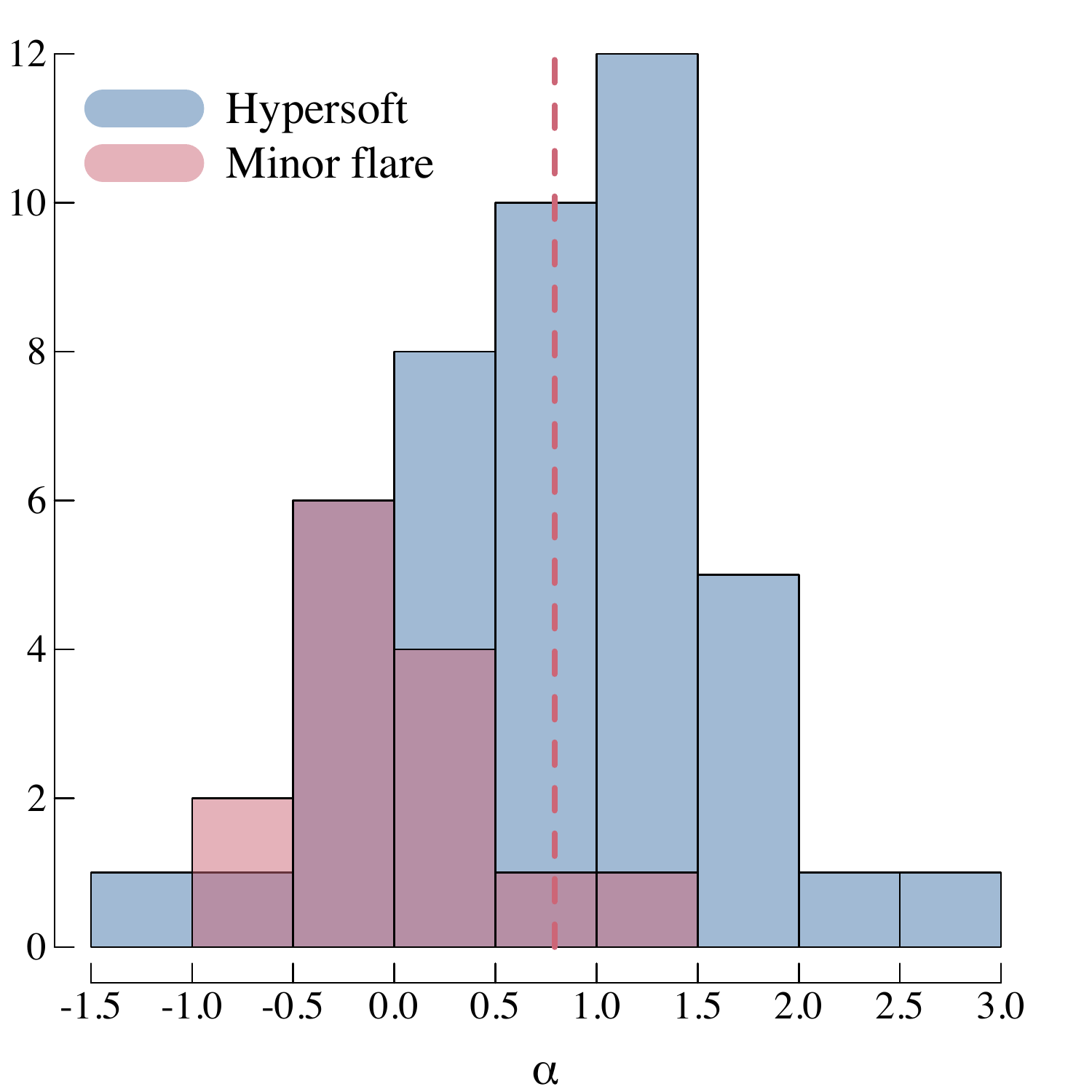}
 \caption{A histogram of the radio spectral index using RATAN-600 data from 4.6 GHz to 11.2 GHz during the hypersoft state (blue) and during the minor radio flaring state (red). The radio spectrum during the hypersoft state is mostly inverted, optically thick, while in the minor flaring state it is flatter or optically thin. The spectral index during the radio spectrum shown in Fig. \ref{radio_spec} is marked as a dotted line.}
 \label{radio_spi}
\end{figure}

During the hypersoft state the simultaneously measured radio spectrum from 4.6 GHz to 11.2 GHz is mostly inverted, i.e. optically thick ($F_{\nu} \propto \nu^{\alpha}$; $\alpha \sim 1$), while during minor radio flaring it is flatter ($\alpha \sim 0$; Fig. \ref{radio_spi}) or optically thin. On the other hand, when we have quasi-simultaneous observations (within 0.5 days) in 15 GHz and 230 GHz bands, the spectrum is approximately flat during both states. However, during the hypersoft state, even slight changes in the radio fluxes can result in a big change in the spectral index, and depend on strict simultaneity (which we do not have for the 15 GHz and 230 GHz bands) as they vary quite rapidly (see above). Overall, we can state that during the hypersoft state the low radio frequencies are absorbed below $\sim$10 GHz, and during minor radio flaring the absorption frequency moves progressively below 5 GHz, while the radio spectra above 10 GHz remains approximately flat in both states.    

\begin{figure}
 \centering
 \includegraphics[width=\linewidth]{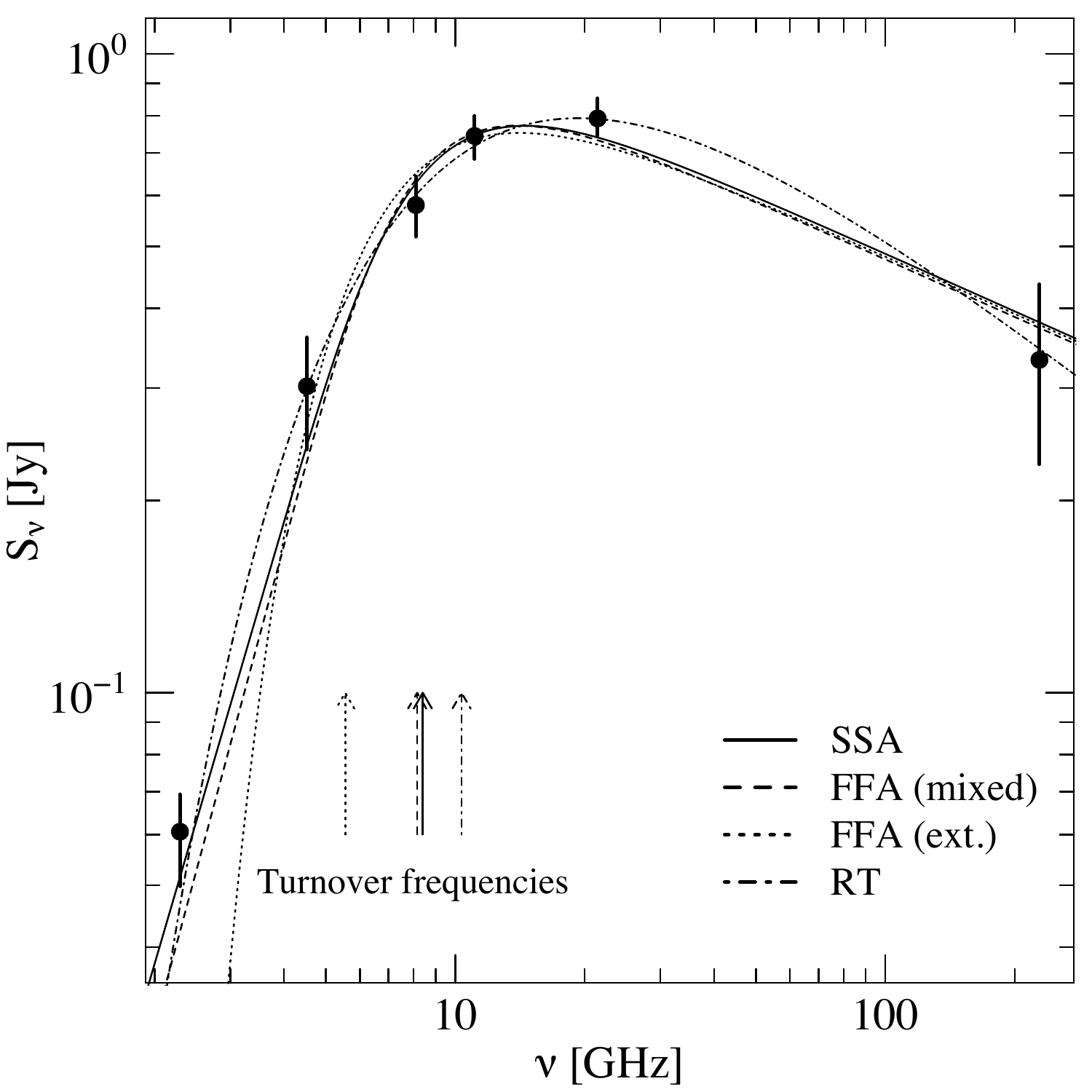}
 \caption{Quasi-simultaneous radio spectrum from RATAN-600 (2--22 GHz; the data in all bands was taken simultaneously on MJD 57631.8) and SMA (230 GHz; the data was taken on MJD 57631.2). The data point for the SMA is an average over the 1.5-hour exposure and the error bars mark the standard deviation of the measurements during the exposure. The spectrum is fitted with several models (SSA: Synchroton self-absorption, FFA: free-free absorption, RT: Razin-Tsytovitch effect). See text for more details.}
 \label{radio_spec}
\end{figure}

The spectral curvature is evident from one example of a quasi-simultaneous (0.6-day difference between RATAN-600 and SMA observations) radio spectrum shown in Fig. \ref{radio_spec} that was taken during the beginning of minor radio flaring in the 2016 outburst episode (MJD 57631.2--57631.8). The spectrum can be adequately fitted by a model of the form \citep{vanderlaan66,gregory74}:

\begin{equation}
S_{\nu} = \frac{S_{0}}{1-\mathrm{exp}(-1)} \, \Bigg( \frac{\nu}{\nu_{0}} \Bigg)^{\alpha_{thick}} \Bigg( 1 - \mathrm{exp} \Bigg[ - \Bigg( \frac{\nu}{\nu_{0}} \Bigg)^{\alpha_{thin}-\alpha_{thick}} \Bigg] \Bigg) \times f(\nu),
 \label{rspecmodel} 
\end{equation} 

\noindent where $S_{0}$ and $\nu_{0}$ are the flux density and the frequency at which $\tau=1$, $\alpha_{thin}$ and $\alpha_{thick}$ correspond to an optically thin and thick spectral index below and above $\nu_{0}$, respectively, and $f(\nu)$ depends on the absorption mechanism. See more discussion in Section \ref{radio_origin}.   

\subsection{X-ray properties}

In Fig. \ref{overview} we see that the \swiftbat\/ daily lightcurve follows roughly the evolution of the radio emission, in a sense that during the hypersoft state the hard X-ray flux and the radio flux density reach their minima and they rise and decay in unison (on a long-term basis from days to weeks) during flaring episodes. This correlation has been established already in \citet{mccollough99}. On the other hand, the soft X-ray flux is at its maximum during the hypersoft state (\maxi\/ 2--10 keV mean flux of 1 cts s$^{-1}$ cm$^{-2}$, and flares up to 2 cts s$^{-1}$ cm$^{-2}$, as compared to 0.4 cts s$^{-1}$ cm$^{-2}$ in the low/hard X-ray state). The soft X-rays are also roughly anticorrelated with the hard X-rays during the hypersoft state, but a robust measure of anticorrelation is difficult to attain because of the negligible \swiftbat\/ flux.          

\subsubsection{X-ray spectrum}

\begin{figure*}
\centering
\begin{subfigure}{.49\textwidth}
  \centering
  \includegraphics[width=\linewidth]{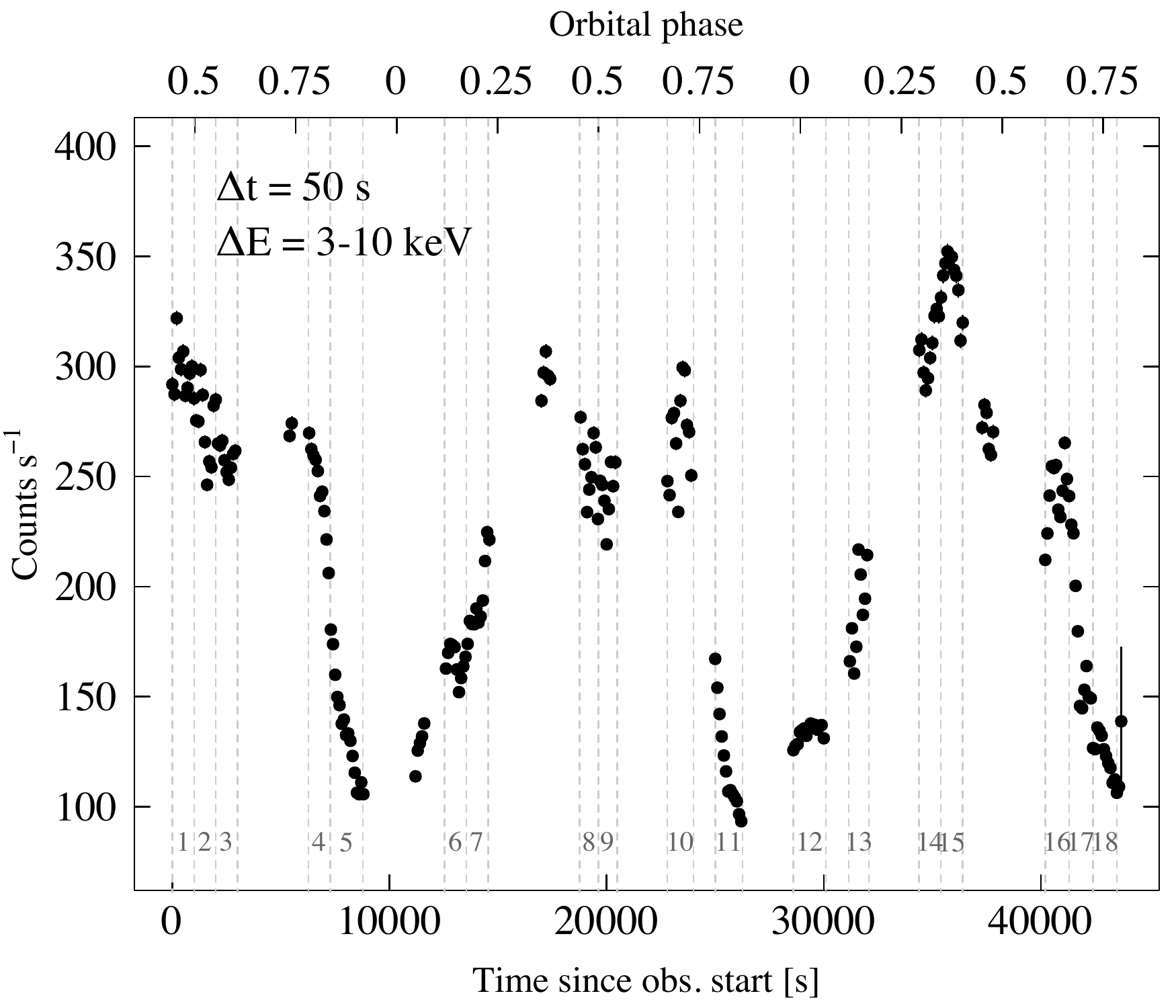}
\end{subfigure}
\begin{subfigure}{.49\textwidth}
  \centering
  \includegraphics[width=\linewidth]{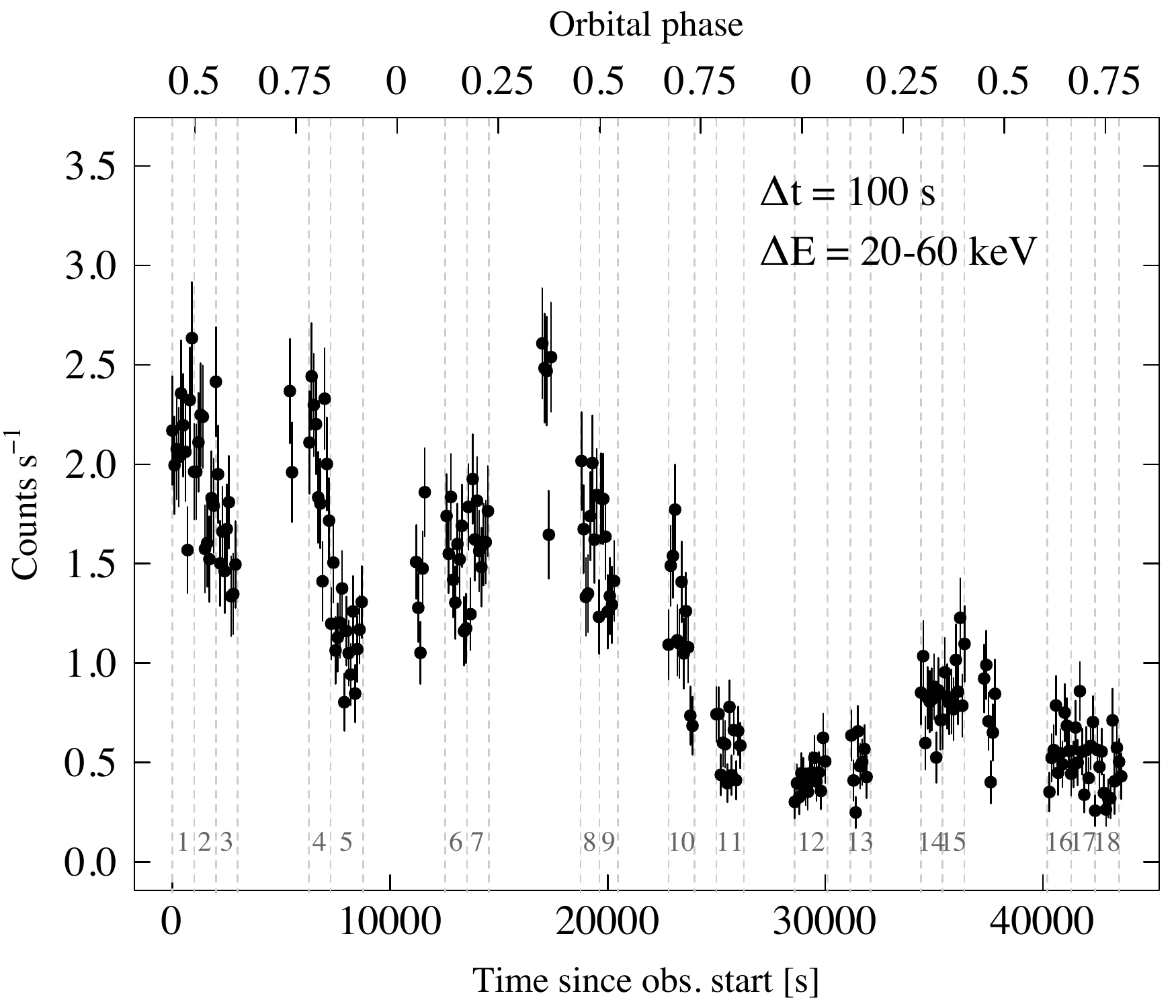}
\end{subfigure}
\caption{\textit{Left:} \nustar\/ count rate in the band 3--10 keV during the whole pointing spanning $\sim$2.5 orbits (orbital phase is labelled on top of the panels). This spectral band consists almost solely of the blackbody emission. \textit{Right:} \nustar\/ countrate in the band 20--60 keV. This spectral band consists of only the power law emission. The dashed lines demarcate the $\sim$1 ksec \nustar\/ spectra that are shown in Fig. \ref{nustar_pha}.}
\label{nustar_lc}
\end{figure*}

\begin{figure*}
  \centering
  \includegraphics[width=\linewidth]{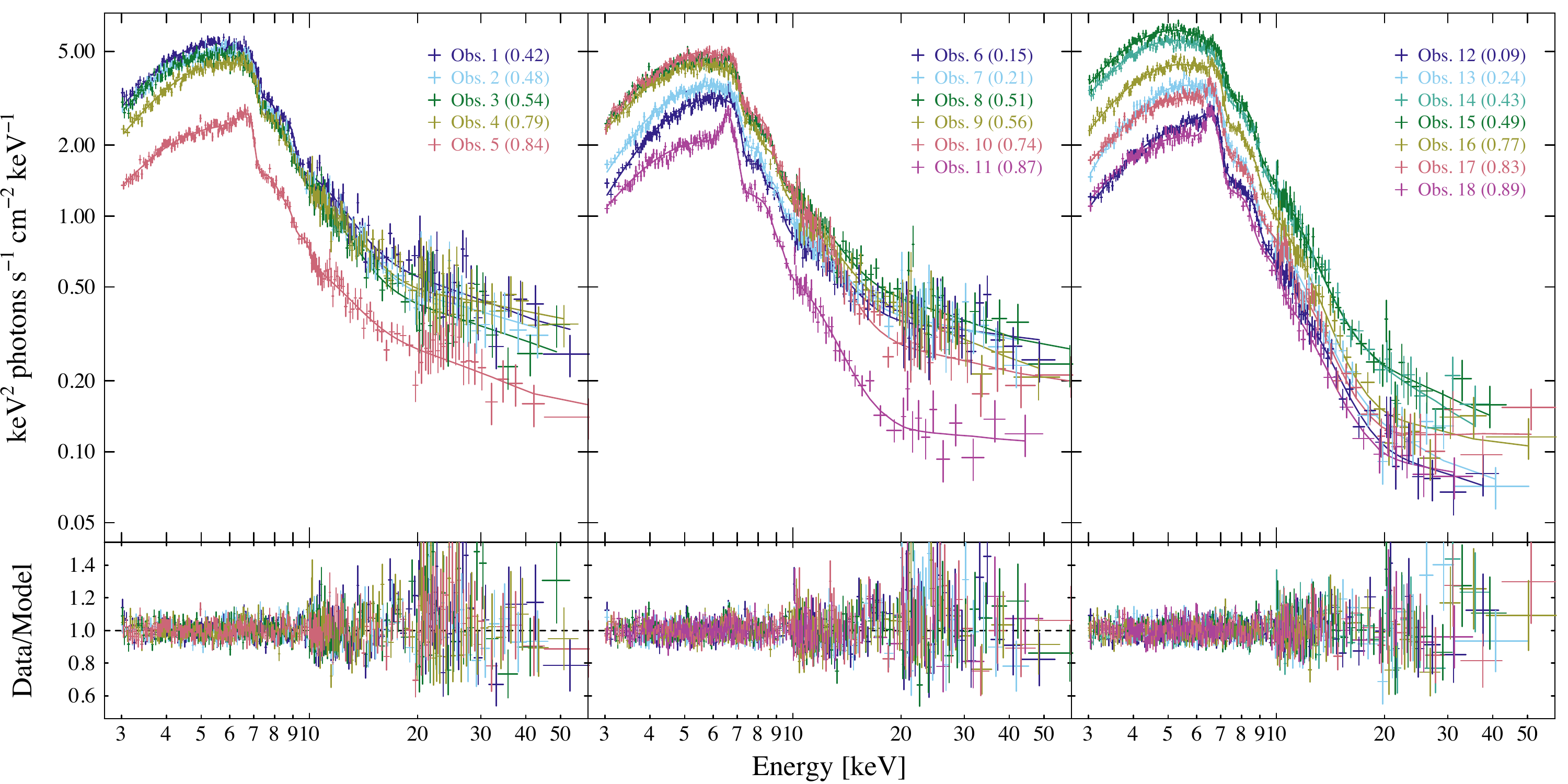}
  \caption{\nustar\/ spectra with $\sim$1 ksec exposure through the whole pointing spanning $\sim$2.5 orbits. Orbital phase is indicated in parenthesis (left panel: 1st half orbit, middle panel: 2nd orbit, right panel: 3rd orbit). The spectra are fitted with a model consisting of absorbed blackbody, Comptonization and gaussian components multiplied by three absorption edges from neutral and ionized iron. The model residuals are in the bottom panels. The plotted spectra are from FMPA for clarity, but the models are fitted to both detectors simultaneously.}
  \label{nustar_pha}
\end{figure*}

As has been found previously, the X-ray spectrum during the hypersoft state is soft; the dominating component being blackbody-like in addition to a weak power law tail \citep{koljonen10,smale93,beckmann07,szostek05}. In these previous works, the soft component has been successfully fitted with a one-temperature blackbody, multicolor blackbody, or heavily thermalized Comptonization model (resembling a blackbody), and the hard component either as Compton upscattered emission from the soft seed photons or a power law model.   

\begin{figure}
  \centering
  \includegraphics[width=0.78\linewidth]{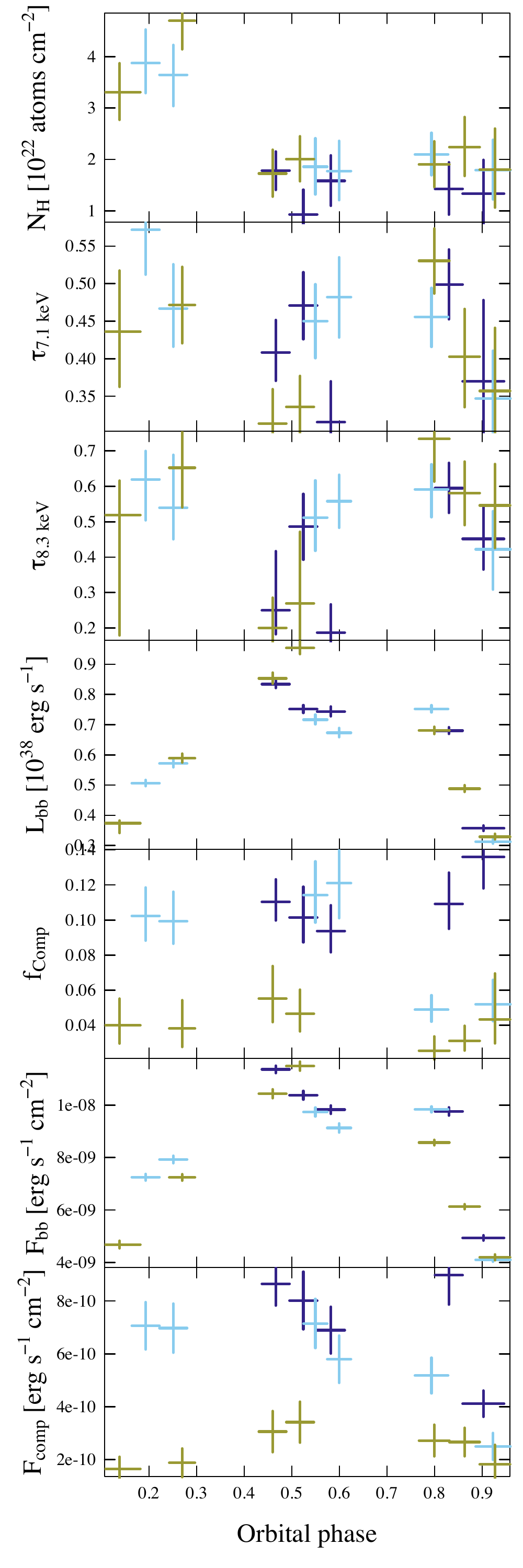}
  \caption{X-ray model parameter evolution over the orbital phase using the \nustar\/ data. The different colors mark the different orbits during the $\sim$20 ksec exposure: blue corresponds to the 1st, light blue to the 2nd and green to the 3rd orbit. From top to bottom: the hydrogen column density, the optical depths of the neutral and helium-like iron, the bolometric blackbody luminosity assuming a distance of 7.4 kpc, the fraction of Compton upscattered seed photons, and the unabsorbed blackbody and Comptonized fluxes calculated from the \nustar\/ energy band 3--79 keV.}
  \label{nustar_params}
\end{figure}

Thus, we started fitting the orbital \nustar\/ spectra ($\sim$1 ksec; see Fig. \ref{nustar_lc} and Fig. \ref{nustar_pha}) from both detectors simultaneously with a model consisting of a blackbody component (\textsc{bbody}) convolved with a Comptonization model \citep[\textsc{simpl};][]{steiner09} and attenuated by interstellar absorption (\textsc{phabs}). This would indicate a scenario where the soft photons are heavily thermalized to one-temperature plasma e.g. by the interaction of the stellar wind with the accretion disk \citep{zdziarski10}. In addition, we add an iron line complex (fitted with a single gaussian line, but likely consisting of neutral and several highly-ionized iron emission lines; see \citealt{paerels00}) and absorption edges of neutral and highly-ionized iron. We considered also a model where the hard X-ray tail was fitted with a power law in case it is not connected to the blackbody component, e.g. in a situation where the hard X-rays are upscattered infrared photons from the jet or optically thin synchrotron emission from the jet. Both models resulted in acceptable fits ($\chi^{2}_{\mathrm{red}}$=0.9--1.4), with the difference that the power law model resulted in a higher absorption column (about twice), as the soft photons are not removed by the Compton scattering. The parameters of the orbital fits with the Comptonization component are plotted in Fig. \ref{nustar_params}.   

For both power law and Comptonization models, the blackbody temperature stayed constant at $kT_{\mathrm{bb}}=1.35\pm0.05$ keV for the whole \nustar\/ observation. The blackbody luminosity is orbitally modulated and peaks at $L_{\mathrm{bb}}=10^{38}$ erg/s (for a distance of 7.4 kpc) during orbital phase 0.5. Alternatively, we fitted the thermal component with a multicolor disk blackbody model (\textsc{ezdiskbb}) with equally good fit quality as the blackbody model, but with increased absorption (two to three times the blackbody model) and a higher temperature ($kT_{\mathrm{bb}}=1.7$ keV).

The column density of the absorption component is orbitally modulated exhibiting highest values, $N_{\mathrm{H}}\sim4\times10^{22} \, \mathrm{cm}^{-2}$, during orbital phase 0.1--0.3 (likely representing local absorption) and otherwise $N_{\mathrm{H}}$ = 1.5--2.0$\times10^{22} \, \mathrm{cm}^{-2}$, which is consistent with the value of the interstellar 21 cm absorption component to the direction of Cyg X-3: $N_{\mathrm{H}} \sim$ 1.6$\times10^{22} \, \mathrm{cm}^{-2}$ \citep{dickey90}. Thus, in order to allow models with a multicolor disk blackbody and/or a power law component, it is necessary to explain the increased, constant absorption in addition to the interstellar component.

When using the power law model, the photon power law index stayed approximately constant at $\Gamma=2.5\pm0.3$. In the case of the Comptonization model, the fraction of seed photons scattering from the electron population is $f_{\mathrm{Comp}}\sim0.1$ in the first two orbits, but drops to half of that during the end of the second orbit (during 3--4 ksec time frame) reflecting the drop in the Comptonized flux. This can be also seen as a drop in the hard X-ray flux in Fig. \ref{nustar_lc}. It is notable that at the same time the radio emission (at 4.6 GHz and 15 GHz) drops from $\sim$600 mJy to $\sim$100 mJy (Fig. \ref{overview}). 

The absorption edges are found to be at energies $E_{\mathrm{Fe/N}} = 7.1-7.2$ keV (neutral iron), $E_{\mathrm{Fe/He}} = 8.2-8.5$ keV and $E_{\mathrm{Fe/H}} = 9-10.5$ keV (highly-ionized iron; helium- and hydrogen-like, respectively). The helium-like edge can be fit better with a smeared edge model (\textsc{smedge}) with its width fixed to $\tau=0.5$ keV, most likely due to the multiple components from the edge profile \citep{kallman04}. There are hints that the optical depth of the absorption edges is reduced at phase 0.5, especially in the case of the third orbit.

The gaussian line is evident in the data only at orbital phase $\sim$0.8--0.1 (Fig. \ref{nustar_pha}). This is consistent with what was found in \citet{vilhu09}, where the iron line flux maximum was attributed close to phase 0.0. They speculated that the orbital phasing of the line photons is due to electron scattering and broad line absorption in the vicinity of the compact object. The centroid energies of the gaussian line are found in the range of $E_{\mathrm{Fe}} = 6.6-6.7$ keV marking the influence of the highly-ionized iron in the iron complex and consistent with the detection of the highly-ionized iron edges. The widths of the gaussian line range between $\sigma_{\mathrm{Fe}} = 0.2-0.4$ keV and the line fluxes between $F_{\mathrm{Fe}} = 0.010-0.016$ photons s$^{-1}$ cm$^{-2}$. 

\begin{figure}
  \centering
  \includegraphics[width=\linewidth]{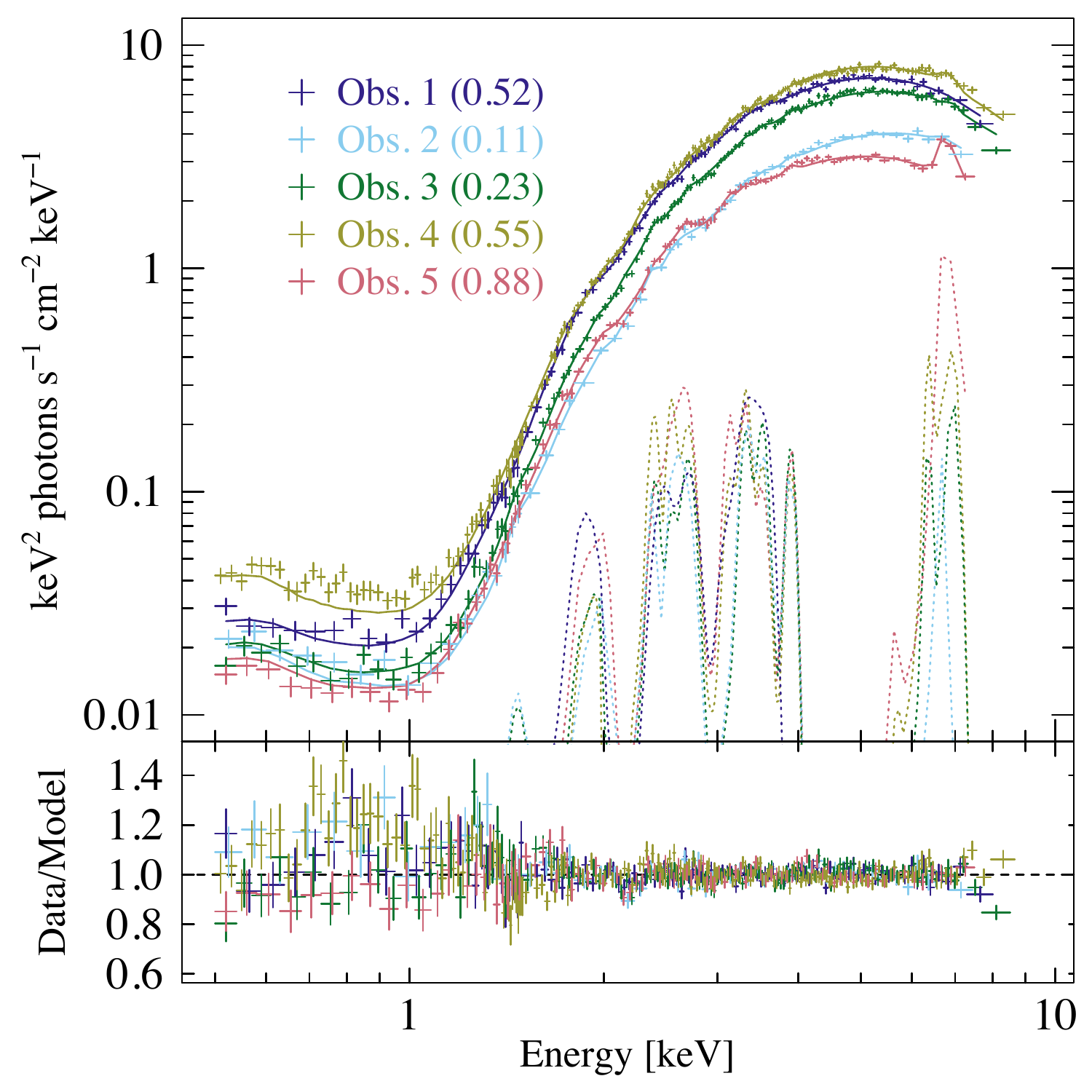}
  \caption{\swiftxrt\/ spectra with $\sim$1 ksec exposure from 2016 (obs. 1--2) and 2017 outbursts (obs. 3--5). Orbital phase is indicated in parenthesis. The spectra have been fitted with a model consisting of an absorbed blackbody modified by a template of absorption/emission lines and radiative recombination continua (dotted lines). The model residuals are in the bottom panel.}
  \label{swift}
\end{figure}

\begin{figure}
  \centering
  \includegraphics[width=0.78\linewidth]{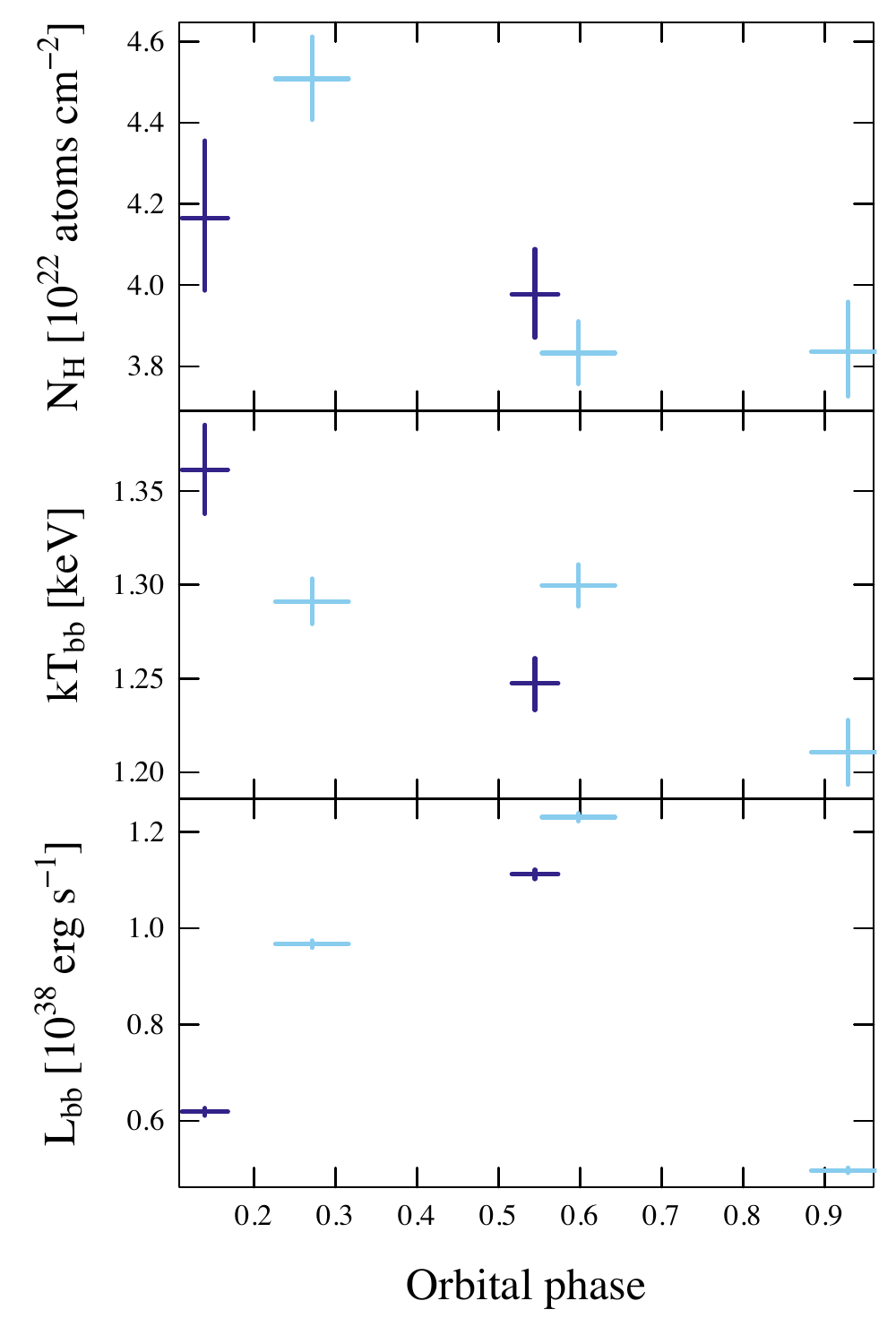}
  \caption{X-ray model parameter evolution over the orbital phase using the \swift\/ data. The different colors mark the different times: blue corresponds to the observations taken in 2016 and light blue in 2017. From top to bottom: the hydrogen column density, the temperature of the blackbody and the bolometric blackbody luminosity assuming a distance of 7.4 kpc.}
  \label{swift_params}
\end{figure}

The \swift\/ spectra (see Fig. \ref{swift}) can be adequately ($\chi^{2}_{\mathrm{red}}$=1.2--1.7) fitted with an absorbed blackbody model modified by a template of emission/absorption lines and radiative recombination continua \citep[][McCollough et al., in prep.]{savolainen12,koljonen13}. The best fit parameters are plotted in Fig. \ref{swift_params}. All phases have $N_{\mathrm{H}}$ $\sim4\times10^{22} \,  \mathrm{cm}^{-2}$ (however, if $N_{\mathrm{H}}$ is frozen to $4\times10^{22} \, \mathrm{cm}^{-2}$ in fitting the \nustar\/ spectra, it does not result in acceptable fit outside phase range 0.1--0.3), with slightly higher absorption ($N_{\mathrm{H}}\sim4.5\times10^{22} \, \mathrm{cm}^{-2}$) during phase 0.2--0.3 similar to (but not in proportion to) the \nustar\/ spectra. Thus, it is possible that the \swift\/ spectra occurring in the hypersoft state present more absorption than during the decay of the minor radio flare that coincided with the \nustar\/ pointing. The emission/absorption lines and radiative recombination continua arise from photoionization of the stellar wind by the intense X-ray radiation \citep{paerels00,szostek08a}. However, a gaussian line component to model the iron line complex is only statistically needed for the last pointing in 2017 with a centroid energy $E_{\mathrm{Fe}}=6.8$ keV, width $\sigma_{\mathrm{Fe}} = 0.2$ keV and a flux $F_{\mathrm{Fe}} = 0.02$ photons s$^{-1}$ cm$^{-2}$. Similar to the \nustar\/ spectra, the temperature of the blackbody model is around $kT_{\mathrm{bb}}=1.3$ keV, except for the last pointing where it has dropped to $kT_{\mathrm{bb}}=1.2$ keV. The effect is clearly visible in the spectrum (Fig. \ref{swift}) as a reduced flux over 4 keV, and cannot be attributed to a change in the absorption profile. As the 2017 \swift\/ pointings occurred at the same date in a subsequent fashion, we can state that the X-ray emitting surface cooled from 1.3 keV to 1.2 keV during a timeframe of 1.3--1.6 hours. Similar to the \nustar\/ results, the blackbody flux is orbitally modulated with a slightly higher maximum of $L_{\mathrm{bb}}=1.2\times10^{38}$ erg/s as compared to the \nustar\/ spectra during phase 0.5. It is interesting to note that the blackbody flux in the last pointing in 2017 does not present unusually low flux with respect to the orbital variation, thus the cooling of the blackbody temperature could imply a rapid change in the blackbody radius.       

\subsubsection{X-ray lightcurve} \label{xlc}

\begin{figure}
  \centering
  \includegraphics[width=\linewidth]{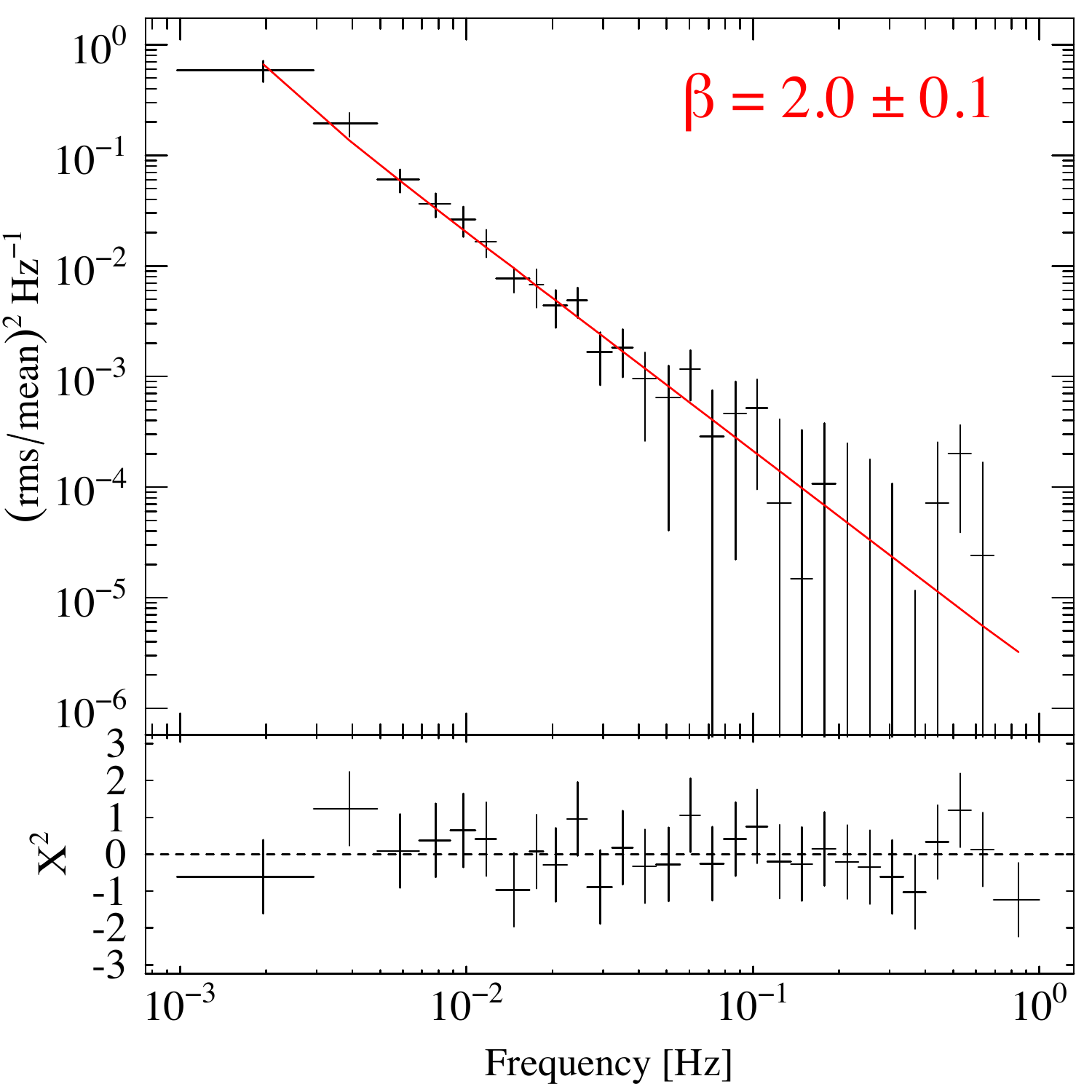}
  \caption{\nustar\/ averaged cospectrum fitted with a powerlaw model.}
  \label{nustar_psd}
\end{figure}

We construct the averaged cospectrum (the real part of the cross spectrum, see Section \ref{nustar_red}) from the \nustar\/ 0.5 s time resolution lightcurves (Fig. \ref{nustar_psd}). The cospectrum technique is used to mitigate the instrumental dead noise in estimating the periodogram of the source. As has been found out previously in \citet{axelsson09,koljonen11}, the PDS of Cyg X-3 can be described as a power law with an index close to $\beta=-2$, with hardly any other structure. In addition, there seems to be no power above 1 Hz. Thus, we fitted the averaged cospectrum with a power law model. Similar to previous results, we find that the power law index is $\beta=-2.0\pm0.1$. 

\section{Discussion} \label{discussion}

To summarize the above observational and modeling results of the hypersoft state we found the following: 

1) The X-ray spectrum can be successfully modelled phenomenologically by using a simple model consisting of a single-temperature blackbody and a power law component. The power law component could arise from the Compton upscattering of the 1.3 keV blackbody photons from a population of high energy electrons that are moving or changing the subtended angle as viewed from the seed photon population in ksec timescales. Alternatively, it could be the optically thin jet emission spanning the X-ray band, or Compton upscattered infrared photons from the jet. Both models resulted in acceptable fits with the only difference in the amount of absorption; close to the interstellar value in the former and twice the interstellar value in the latter, with the exception in orbital phases 0.0--0.3, where local absorption is present in both cases. The single-temperature blackbody model can be also exchanged to multicolor blackbody component, but again with the expense of introducing a higher absorption column (a factor of 2--3) and a higher blackbody temperature.   

2) There are strong absorption edges present in the X-ray spectra indicating screens of neutral and ionized iron in the line-of-sight, and ionized iron line that is visible in the X-ray spectra only at orbital phases 0.8--0.0, most likely occurring in a photoionized region close to the compact object.

3) The X-ray cospectrum (a proxy for PDS) is similar to what is observed in the PDS of every other spectral state: power law with an index of $\beta=-2.0$. 

4) The radio spectrum is approximately flat (or slightly inverted) from $\sim$10 GHz to 230 GHz (also during the $\sim$ 1 Jy radio flare), and present the lowest fluxes in the source evolution down to 1 mJy level with the radio variability being at its maximum. During the hypersoft state and beginning of a minor radio flare, the radio fluxes below 10 GHz are absorbed with a relatively high index indicative of free-free absorption, synchrotron self-absorption or a Razin-Tsytovitch effect. The radio emission was also found to be coupled to the hard X-ray emission in orbital timescales. 

5) Significant gamma-ray emission usually occurs during the hypersoft state. In particular, at times corresponding to the onset of the hypersoft state, the onset or during the minor radio flare and during the onset of the major radio flare. All of these were observed in the 2016 and 2017 data sets.   

\subsection{Origin of the radio emission} \label{radio_origin}

The radio spectrum has been previously speculated to arise as a synchrotron emission from expanding plasmons \citep[e.g.][]{marti92,fender97} or shocks in the jet \citep{lindfors07,millerjones09}. Following \citet{gregory74} and \citet{millerjones04}, this emission is thought to be absorbed either by synchrotron self-absorption, free-free absorption by thermal plasma uniformly mixed with or surrounding the synchrotron emitting particles, or a suppression of synchrotron radiation by a dispersive medium in a so-called Razin-Tsytovitch effect \citep{hornby66,simon69}. Of course, a mixture of these processes could also take place, but in the following, we consider them separately. The low-frequency cut-off could also arise from induced Compton scattering \citep{sincell94} or a low-energy cut-off in the electron spectrum. However, in the latter case, the slope below the turnover would be $\alpha_{thick}\sim0.3$, which is too shallow to account for the observations.      

In the above, it was speculated that the radio emission during the hypersoft state is likely similar to what was observed during the beginning of the minor radio flare. We have fitted the minor flare radio spectrum (Fig. \ref{radio_spec}) with a non-thermal model absorbed by the above-mentioned absorption processes according to Eq. \ref{rspecmodel}. For the synchrotron self-absorption ($f(\nu)=1$), setting $\alpha_{thick}=2.5$ and $\alpha_{thin}=-0.3$ produced an acceptable fit to the spectrum. Previously, the synchrotron self-absorption has been deemed as sufficient to explain fully the absorption of the radio flux densities in the shock-in-jet model during major and minor radio flares \citep{millerjones09}. The synchrotron self-absorption turnover frequency is given approximately as \citep[e.g.][]{kellermann81},

\begin{equation}
\nu_{0} \sim f(\gamma) B^{1/5} S_{0}^{2/5} \theta^{-4/5} \, \mathrm{GHz},
\end{equation}
     
\noindent in the case of uniform source and power law distribution of particles with an index $\gamma$. The function f($\gamma$) does not depend strongly on $\gamma$, and it is f($\gamma$)$\sim$8 if $\gamma$=2. With the flux density of S$_{0}$ = 0.7 Jy during the peak of the minor radio flare, and a VLBI source angular size $\theta\sim1$ mas \citep[][also corresponds to the radio photosphere size at 15 GHz from Eq. \ref{photosphere}]{egron17}, the magnetic field strength should be 6--100 G. If assuming that the same spectral shape applies to the hypersoft state where S$_{0}\sim$ 0.01 Jy, the magnetic field strength should be much larger ($>3\times10^{4}$ G). These values are much larger than estimated previously \citep[$\sim$0.1 G;][]{gregory74,fender97,millerjones04}. However, as the radio images of Cyg X-3 are scatter-broadened, the true source size is difficult to determine, and therefore the magnetic field strength could be lower. 

Synchrotron self-absorption becomes important for sources with brightness temperatures $T_{b} > m_{e}c^{2}/3k \sim 2\times10^{9}$ K. The brightness temperature of Cyg X-3 can be approximated from the duration of a flare and its flux change \citep{ogley01}: 

\begin{equation}
T_{b} \geq 9.66 \times 10^{10} \frac{\Delta S D^{2}}{\nu^{2} \Delta t^{2}} \, \mathrm{K},
\end{equation}

\noindent where $S$ is in units of mJy, $D$ in kpc, $\nu$ in GHz and $t$ in minutes. During hypersoft state (see Fig. \ref{radio_detail}) the flare duration $\Delta t \sim0.1$ day (140 minutes), flux change of $\Delta S = 10$ mJy at $\nu = 15$ GHz and a distance of $D=7.4$ kpc implies brightness temperature of $T_{b} \geq 1\times10^{7}$ K (at the minor flare peak it increases to $T_{b} \geq 3\times10^{8}$ K). Thus, this value is too low for the synchrotron self-absorption to take place unless the source size is much smaller than implied by the flare duration. Therefore, we cannot rule out the synchrotron self-absorption as a possible mechanism for the observed absorption, but we deem it unlikely, as such a high synchrotron self-absorption turnover frequency have not been observed from any other flat radio spectrum XRB, and if observed they have been found to lie below 1 GHz. The low brightness temperature would also exclude induced Compton scattering as causing the turnover. 

The mixed free-free absorption has been proposed previously to best represent the absorbed radio spectrum \citep{gregory74} and to explain the decreasing opacities in a series of radio flares \citep{fender97}. Here, the thermal plasma surrounding the compact object is thought to be entrained in the jet during the launching of the jet after the quenched period. For the mixed free-free absorption ($f(\nu)=1$), setting $\alpha_{thick}=2.1$ and $\alpha_{thin}=-0.3$ in Eq. \ref{rspecmodel} produced also an acceptable fit to the spectrum (Fig. \ref{radio_spec}). 

The external free-free absorption has been discussed in \citet{fender95,waltman96} to be the cause of the absorption as an absorbing screen of thermal plasma likely from the WR wind. The radio spectrum can be fitted with the external free-free absorption by setting $\alpha_{thick}=\alpha_{thin}=-0.3$ and $f(\nu)=\mathrm{exp}[-(\nu/\nu_{0})^{-2.1}]/\mathrm{exp}(-1)$ in Eq. \ref{rspecmodel}. However, at low frequencies, the model seems to drop too fast in order to account for the data point at 2.2 GHz.    

The Razin-Tsytovitch effect causes the suppression of synchrotron radiation and it occurs when the relativistic electrons are surrounded by plasma. The refractive index of the plasma reduces the Lorentz factors of the electrons which result in a low-frequency cut-off. The frequency of this cut-off is given approximately by $\nu_{\mathrm{RT}} = 20 n_{e}/B$ Hz, where the magnetic field strength $B$ is in units of Gauss and the electron density $n_{e}$ in cm$^{-3}$. Thus, the synchrotron emission decreases exponentially below the cut-off frequency, and the spectrum can be approximated using Eq. \ref{rspecmodel} with $\alpha_{thick}$ = $\alpha_{thin} \sim -0.5$, and $f(\nu)=\mathrm{exp}[-(\nu_{\mathrm{RT}}/\nu_{0})]/\mathrm{exp}(-1)$. With 10 GHz cut-off frequency, $n_{e}/B$ = 5 $\times$ 10$^{8}$. We can estimate the electron number density using the same stellar wind parameters as in Eq. \ref{photosphere} and assume that the radio emission is coming at 1 mas ($r \sim 6 \times 10^{13}$ cm): 

\begin{equation}
n_{e} \approx \frac{\dot{M}\gamma}{4 \pi r^{2} v_{\infty} m_{H} \mu} = 2 \times 10^{7} \, \mathrm{cm}^{-3}. 
\end{equation}  

Thus, the magnetic field has to be in excess of 0.07 G in order to for the Razin-Tsytovitch effect to take place, which is a reasonable assumption based on the previous estimates.

From the above, we conclude that the synchrotron emission absorbed by the mixed free-free absorption or the Razin-Tsytovitch effect represents best the radio spectrum taken at the beginning of the minor radio flare. Due to the similar spectral indices obtained at frequencies 4.6--11.2 GHz, this is likely true during the hypersoft state as well.        
 
\subsection{Jet quenching}

During the hypersoft state, there is clearly radio emission left and opacity is still a factor. However, the radio flux density at least up to 230 GHz is greatly reduced by as much as two orders of magnitude from the baseline/non-flaring state level (similar quenching factors have been observed from other XRBs, e.g. \citealt{fender99,russell11,rushton16}). The spectrum of the quenched radio emission is not optically thin which would be expected from residual radio lobes. Instead, the spectrum is optically thick with an index close to $\sim$1, and likely caused by free-free absorption or Razin-Tsytovitch effect (see above). \citet{waltman96} discussed that the reduced radio flux densities are unlikely to be absorbed by an enhancement of the WR stellar wind, as the onset of quenched emission should appear later at lower frequencies if the enhanced stellar wind density overtakes the radio photospheres. However, the process of quenching of the radio emission does not appear to have any lag depending on the frequency, thus they favored that most likely the mass injection rate to the jet is quenched. 

\citet{fender97} suggested that during the radio quenched period the mass-loss rate of the WR star increases resulting in an increase of the thermal electron density at the location of the jet, which effectively quenches the jet by a mechanism unknown. At the same time, the infrared emission of the stellar wind is increased due to a higher ion and electron densities, as well as the X-ray emission due to increased mass accretion rate. \citet{mccollough10} show that during the soft X-ray state there is a clear brightening (0.3 mag) in the infrared emission in all 2MASS bands. Assuming that the infrared flux comes as a whole from the stellar wind, 0.3 mag brightening correspond to a factor of 1.3 increase in flux, which corresponds to a factor of 1.4 increase in mass-loss rate assuming that $S_{\nu} \approx \dot{m}^{4/3}$. The infrared flux reaches its maximum during the hypersoft state and the following radio flare \citep[][their Fig. 2]{mccollough10}. However, it can be also seen that the actual state change has started earlier, 10--15 days before the rise in the infrared, when soft X-rays started to increase and hard X-rays to decrease. If the X-ray spectrum gets softer, the opacity of the wind increases, since there are line/edge absorption processes operating that are not important at the hard X-ray energies. This results in the increase of the wind temperature and subsequently higher infrared radiation. Thus, it seems likely that changes in the accretion flow are driving the quenching rather than changes in the wind mass-loss rate.

Whether the continuum emission comes from the relativistic jet during the hypersoft state is not clear. We cannot exclude a possibility that a weak, self-absorbed jet is still present in the system. However, in the soft states of XRBs, strong disk winds are seen \citep{neilsen09,ponti12}, and these may collide with the WR stellar wind to produce radio emission at the shock front. Systems with colliding stellar winds produce radio emission with luminosities 10$^{29}$--10$^{30}$ erg/s \citep{debecker13}, that correspond to a flux density of 1--5 mJy for a distance 7.4 kpc at 5 GHz. The X-ray source could heat the wind even further and lead to a larger flux in free-free emission from the wind.  

Furthermore, the correlation of the hard X-ray emission to the radio emission during the hypersoft state hints a coupling between these emission regions. The coupling can be explained e.g. by a model of \citet{meier01}, where the geometrical thickness of the accretion flow is related to its ability to hold a strong poloidal magnetic field. If the hard X-rays are connected either to the base of the jet or the thick accretion flow, then the jet quenching would naturally be correlated with the lack of the hard X-ray emission.     

\subsection{Origin of the X-ray emission}

\citet{zdziarski10} showed that the Cyg X-3 spectral energy distribution and the high-frequency power spectra in both soft and hard spectral states can be modelled by assuming that the X-ray source is embedded in a Compton-thick, low-temperature plasma. Compton scattering in the plasma results in a redistribution and isotropisation of the photon spectrum, in such a way that the high-frequency variability of the X-rays is suppressed and the high-energy photons are scattered to lower energies towards the equilibrium temperature of the plasma. The necessary optical depth and equilibrium temperature are however much larger ($\tau\sim7$, $kT_{\mathrm{bb}}\sim2.5$ keV) than what would be expected from the stellar wind alone, and thus they speculated that the interaction of the (focused) stellar wind with the accretion disk would produce a bulge around the compact object with the necessary properties. 

The energy spectrum and the cospectrum in the hypersoft state support the scattering scenario. We observe strong thermalisation of the incident spectrum, as the hypersoft spectra can be fitted purely with blackbody model. The temperature of the blackbody is lower ($kT_{\mathrm{bb}}\sim1.3$ keV) as compared to the value in \citet{zdziarski10}, but this obviously depends on the shape of the incident spectrum and the local absorption external to the scattering cloud. 

The X-ray cospectrum is a featureless power law spectrum with an index of $\beta=-2.0$, similar to what has been observed previously from the PDS of all accretion states of Cyg X-3 \citep{axelsson09,koljonen11}. This differs what is usually seen from XRBs, that produce flicker noise spectra with flatter index of $\beta\sim-1$ in the soft state \citep[e.g.][]{gilfanov05}. The power law PDS with an index $\beta = -2$ arises usually from some kind of red noise process, where high-frequency variations are suppressed \citep{kylafis87}. Possibly, as the system is enshrouded in a dense stellar wind, the X-ray scattering in the stellar wind will result in further suppressing the high-frequency variations mimicking a red noise process. The similarity of the X-ray timing properties in all accretion states implies that the density of the stellar wind stays above a certain density threshold that produces enough X-ray scattering at all times, and the changes in the energy spectrum are due to changes in the accretion process.  

Similar X-ray spectral properties as in Cyg X-3 during hypersoft state have been observed in GRO J1655--40 \citep{uttley15,neilsen16,shidatsu16}. \citet{uttley15} found that during the 2005 outburst the source presented suspiciously similar spectra with Cyg X-3 featuring an absorption edge at $E_{\mathrm{Fe/He}} = 8.3$ keV, constant disk temperature of $kT_{\mathrm{bb}}=1.2$ keV and a Compton scattering fraction of $f_{\mathrm{Comp}} = 0.2$. However, the X-ray PDS of GRO J1655--40 does not feature red noise type noise but a flat spectrum with a low-frequency cut-off ranging $\nu_{\mathrm{cut}} = 0.1-10$ Hz. \citet{neilsen16} and \citet{shidatsu16} argued that the origin of the X-ray spectral and timing properties together with an infrared excess could be explained by a Compton-thick disk wind produced by the accretion disk emitting at or above the Eddington limit. In addition, during the hypersoft state in GRO J1655--40 the radio is also quenched \citep{migliari07}. Thus, it could be possible that Cyg X-3 presents a similar state of super-Eddington accretion in the hypersoft state as in GRO J1655--40 with similar symptoms of soft X-ray spectrum, infrared excess, and radio quenching. 

On the other hand, Swift J1753.5--0124 exhibited an unusual spectral state in March--May 2015 with similar properties to the hypersoft state with soft X-ray spectrum and \swiftbat\/ flux consistent with zero \citep{shaw16}. At the same, time radio flux was not detected from the source and upper limits with a quenching factor of $>$25 were obtained \citep{rushton16}. However, the X-ray luminosity was shown to be unusually low for a soft state XRB: approximately 0.6\% of the Eddington luminosity for a distance of 3 kpc and black hole mass 5 $M_{\odot}$ \citep{shaw16}. As the distance is not well established \citep[2--8 kpc;][]{cadollebel07,froning14}, the Eddington luminosity could be as high as 4\%. This indicates that the hypersoft state and jet quenching does not necessarily need Eddington limit accretion rates. In fact, the unabsorbed X-ray luminosity of Cyg X-3 in the hypersoft state ($\sim10^{38}$ erg/s) indicates $\sim$4--8 \% of the Eddington luminosity (which is roughly twice the `normal' Eddington luminosity as most of the accreted mass is helium) for a black hole accretor of 5--10 $M_{\odot}$. It is unlikely that the hypersoft state in Swift J1753.5--0124 would be a result of a Compton thick wind as the X-ray luminosity is very low to drive a strong disk wind, and the inclination is likely not very high \citep[$<55^{\circ}$;][]{froning14}. Instead, a Compton-thick disk atmosphere could scatter the accretion disk emission as speculated in \citet{shaw16}. 

Thus, the common theme in these three sources is surprisingly similar: thermal X-ray spectrum with a very weak or non-existent hard X-ray tail coinciding with a very weak or non-existent radio emission. The spectral similarity suggests that the spectral shape is intrinsic to the hypersoft state and is not due to scattering, as these systems have likely different origins and properties of the intervening medium (disk wind, disk atmosphere, stellar wind). The lack of radio emission in all sources during the hypersoft state suggest that the accretion flow is in a state (not necessarily connected to the Eddington luminosity of the flow) that effectively quenches the jet formation. 
 
\subsection{Origin of the $\gamma$-ray emission}

The detection of $\gamma$-ray emission in Cyg X-3 requires an efficient mechanism of producing relativistic particles. This naturally occurs in shocks. The $\gamma$-ray emission in Cyg X-3 seems to be usually associated with the hypersoft state \citep{tavani09,koljonen10}. A seemingly strong anti-correlation exists between the $\gamma$-ray emission and the hard X-ray emission, and it appears that every local minimum in the hard X-ray lightcurve corresponds to a time when $\gamma$-ray emission is detected from the system. The $\gamma$-ray flares are usually observed during the declining phase as well as the rising phase to/from the hypersoft state \citep{tavani09}. It is still under debate whether the $\gamma$-rays arise from leptonic \citep{dubus10,zdziarski12,piano12} or hadronic processes \citep[e.g.][]{romero03,piano12,sahakyan14}. The leptonic scenario is based on the inverse-Compton scattering of soft stellar photons by the relativistic electrons presumably in the jet that is streaming towards us close to the line-of-sight. The interaction of the jet with the stellar photon field close to the compact object (within ten orbital separations) scatters the photons up to MeV energies. It is also possible that the interaction of the jet with the stellar wind will produce a strong recollimation shock in the jet that can lead to strong dissipation of kinetic energy into $\gamma$-ray and radio emission \citep{yoon15,yoon16}. In addition, clumps in the stellar wind can enhance the production of shocks and subsequently the $\gamma$-ray emission \citep{delacita17}. The hadronic scenario, where the jet would be populated with protons, is based on the proton-proton collisions that occur between the protons in the hadronic jet and the protons in the stellar wind. The collisions produce pions and $\gamma$-rays via the decay of neutral pions.

However, $\gamma$-rays have also been detected outside of major radio flare episodes and have been associated with very brief periods of hard X-ray quenching and minor radio flaring \citep{corbel12,bulgarelli12}. In one case the hard X-ray quenching lasted only $\sim$0.5 days and in the other for a day. Interestingly, $\gamma$-rays are reported to occur also without associated enhancement of the radio emission \citep{williams11} and/or hard X-ray quenching \citep{bodaghee13}. Thus, the jet origin of the $\gamma$-rays is still debatable as both \citet{williams11} and \citet{bodaghee13} present some evidence against the model of inverse-Compton scattering of soft stellar photons by the jet. 

We note briefly here that if Cyg X-3 exhibits strong disk wind similar to GRO J1655--40, its collision with the stellar wind may be strong enough to provide some or all of the $\gamma$-rays from Cyg X-3. Overall, the $\gamma$-ray luminosity is a factor of about 100 larger than that from the colliding wind binary $\eta$ Carinae \citep{abdo10}. The mass loss rates and velocities of the winds of the two systems are quite similar. The density at the shock interface in Cyg X-3 should be much larger, resulting from smaller orbital separation. A detailed study of this possibility is out of the scope of this work, but it appears to be a viable possibility that the disk wind and stellar wind interaction may supply at least some of the $\gamma$-ray power in Cyg X-3.

\subsection{The proposed scenario}

In order to tie together all the above phenomena observed from Cyg X-3, we propose the following scenario, that expands on the ones presented in \citet{fender97} and \citet{vadawale03}. In the former, the thermal electron density in the vicinity of the compact object increases (due to increased mass-loss from the WR companion) and quenches the jet launching mechanism. The increased mass-loss results in the increase of mass accretion rate to the compact object, which eventually leads to a large injection of matter into the jet. Due to the higher thermal electron density, the relativistic particles in the jet are mixed with a high proportion of absorbing thermal electrons producing opacity to the synchrotron emission. As the mass-loss rate decreases with time, the opacity also decreases in the radio flares. The scenario in \citet{vadawale03} is based on the internal shock model \citep{rees78,marscher85,kaiser00}, where the travelling plasmons in the jet are associated with shocks that arise from the differential velocity of the moving jet material. Considering the case of GRS 1915$+$105, they presented a model where during a hard-to-soft state transition unstable accretion flow changes result in discrete, multiple ejections of matter in the jet which collide with the continuous, flat spectrum jet that is present in XRBs during the hard state and produce shocks and episodic radio flares. On the other hand, during the soft state, the jet is quenched, and the reignition of the jet when the source transits from the soft state back to the hard state does not result in shock formation and subsequent radio flaring as there is not material present for the jet to collide with.  

\begin{figure}
  \centering
  \includegraphics[width=\linewidth]{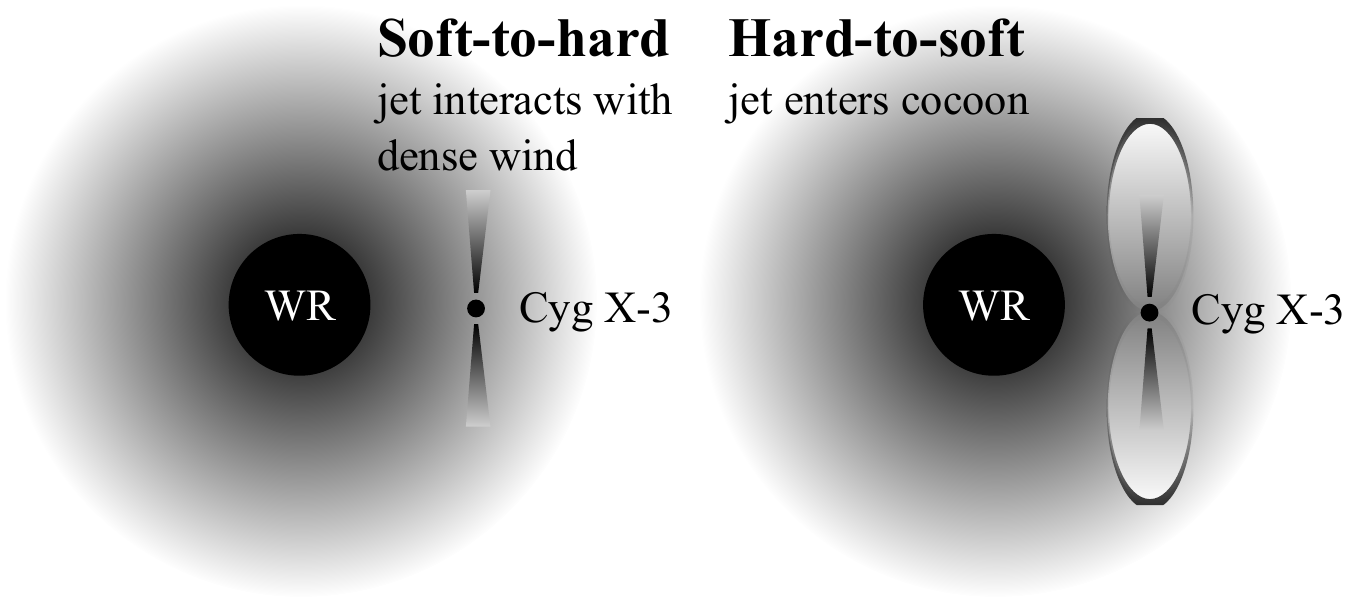}
  \caption{The cartoon of the jet-wind interaction for Cyg X-3. \textit{Left:} In the hypersoft state the jet is turned off, and the stellar wind is allowed to expand freely close to the compact object. When the jet turns back on during the state transition, it encounters a dense medium at rest relative to the jet which leads to efficient shock production and subsequently to strong radio and $\gamma$-ray emission. \textit{Right:} When Cyg X-3 is transiting from the hard state to the hypersoft state, the hard state compact jet has blown out a cocoon in the stellar wind, into which the episodic jet can then expand more freely producing weaker radio emission than when expanding directly into a stellar wind. \textit{Note:} The stellar sizes and the binary separation are not to scale with respect to the jet.}
  \label{cartoon}
\end{figure}

Here, we posit that the work surface for the jet can be either the stellar wind or the continuous, flat spectrum jet. Thus, we develop a scenario where systems with strong stellar winds should see more powerful radio episodes at the transitions from soft states to hard states than at the transitions from hard states to soft states. In the low/hard state, a jet constantly does work against the stellar wind in the system, evacuating a cocoon in a similar fashion as the jets in SS 433 \citep[e.g.][]{fabrika04} and ultra-luminous XRBs \citep[e.g.][]{pakull10}. Possibly, the formation of a strong recollimation shock in the jet \citep{yoon15, yoon16} enhances the radio emission to levels above normal, persistent XRB, that exhibit radio flux densities below 20 mJy with smaller distances \citep{fender00}. Due to changes in the accretion flow, the accretion rate to the compact object increases resulting in higher soft X-ray luminosity and higher infrared luminosity as discussed above. As the source descends to the hypersoft state, the jet quenches after producing episodic shocks similar to other XRBs when transiting to the soft state. When these shocks interact with the stellar wind \citep[or possibly with the clumps in the wind;][]{delacita17}, they produce $\gamma$-ray emission. During the time that the jet stays quenched, the stellar wind is able to fill the cocoon region back in. When the source makes a transition back to the low/hard state, the jet turns back on, and it encounters a medium with the density of the stellar wind. This medium is far denser than the relativistic jet would normally be in and is also essentially at rest relative to the fast jet (i.e. it moves at a speed a factor of about 100 slower). This interaction leads to efficient shock production in the jet and radio and $\gamma$-ray emission \citep{yoon15, yoon16}. The thermal matter of the stellar wind is entrained in the jet producing increased opacity of the synchrotron radiation until the cocoon is blown again by the jet pressure. A cartoon of this paradigm is presented in Fig. \ref{cartoon}.

It has often been stated that the jets turn off in the soft states of black hole XRBs simply because the radio emission turns off. An alternative, in principle, is that the radiative efficiency drops in the jet (e.g. ``dark jets'' in soft states, \citealt{drappeau17}; or Poynting flux dominated jets, e.g. \citealt{sikora05}) or that the jet Lorentz factors become sufficiently large that observers sitting outside the beaming cone simply do not detect flux from them \citep{maccarone05}. The scenario we propose here strengthens the evidence for the idea that the jets really do turn off or become very weak in the soft state. If the jets remained on with a similar jet power as in the hard state, then there would remain an evacuated cavity in the stellar wind from Cyg X-3's donor star even during the soft state, and the radio power upon returning to the hard state would not be dramatically higher than at the hard-to-soft transition. It seems unlikely that a Poynting flux dominated jet could travel through such a large amount of material without depositing energy but a detailed calculation of this scenario does not exist in the literature.
 
In recent years, a controversy has developed surrounding whether it can be demonstrated that black hole spin has an important effect on the jet properties \citep[e.g.][]{narayan12,russell13,koljonen15}. In most cases, the radio luminosity of the jet is used as a proxy for its kinetic power, although in some cases, with careful modelling, the jet's kinetic power can be estimated from the jet-interstellar medium interactions \citep{gallo05,russell07,cooke07,sell15}. The latter approach is frequently used in AGN studies due to concerns about the effects of beaming if core radio fluxes are used. Low-mass XRBs show strong hysteresis effects in their state transition luminosities, with this having been found both for black hole systems \citep{miyamoto95} and neutron star systems \citep{maccarone03}. This means that the hard-to-soft state transitions for the low-mass XRBs, in addition to having denser media for internal shocking than the jets produced at the soft-to-hard transition, they also have higher X-ray luminosities at the transition and likely higher kinetic powers apart from any spin effects. Cyg X-3, like Cyg X-1, does not show hysteresis effects in its state transitions, probably because its outer disk radius, as a wind-fed system, is small enough that it is not a bona fide transient in the X-rays \citep[e.g.][]{smith02}. Thus, differences in the accretion luminosity cannot be used to explain the differences in jet radio luminosity; they must come from either a difference in how accretion power is converted into jet power, or how the jet power is dissipated and converted into radio emission.
 
Our explanation for Cyg X-3's behaviour, as coming from the very dense medium in which the jet kinetic power is dissipated, rather than from the differences in kinetic power itself, thus has implications for all attempts to estimate the kinetic power of transient jets based on radio flux. It is likely that the spread of variation in the densities of ambient media around low-mass XRBs is substantially smaller than the difference between these typical values and the density of the stellar wind around Cyg X-3, but this remains a potential error source not well discussed in the current literature.

\section{Conclusions} \label{conclusions}

We have studied the multiwavelength properties of Cyg X-3 during the hypersoft state that is observed prior to major outburst events and jet ejection episodes. We have shown that the radio/sub-mm emission is diminished by two orders of magnitudes and present the lowest radio flux densities in the source evolution. The radio emission size region is the most compact in the source evolution, however, it is much larger than the orbital separation of the binary. The radio emission in the low-frequencies during the hypersoft state (and in the beginning of radio flares) is likely absorbed by thermal plasma mixed with the non-thermal electron population. We have also suggested that the `residual' radio emission in the hypersoft state could arise from the wind-wind interaction of the binary, and not from the jet. 

We have shown that the broad-band X-ray spectrum is consistent with a thermal, absorbed blackbody emission and a Comptonized component from a population of high energy electrons that are moving or changing the subtended angle as viewed from the seed photon population in ksec timescales. The radio emission was also found to be coupled to the hard X-ray emission at the same timescale. The X-ray spectra are subjected to absorption and electron scattering by the stellar wind, that seems to be manifest in all accretion states. Thus, the accretion state change to/from the hypersoft state is linked to changes in the accretion flow and not in the stellar wind structure.  

Cyg X-3 is an XRB which shows the standard spectral state transition phenomenology of XRBs but which shows a different jet radio power phenomenology. We have shown that a coherent picture of the behavior of the system can be developed if one considers the effects of the stellar wind on the radio emission, and of the jet on the stellar wind density. In the hard state, the jet constantly evacuates a cocoon in the stellar wind, while in the hypersoft state, the wind refills this region, providing a work surface for the jet when the disk returns to the hard state. Our scenario, in particular, provides evidence that the jets actually are quenched in the soft states, rather than becoming radiatively inefficient or travelling with very high Lorentz factors such that the flux outside the beaming cone is strongly deboosted.

\section*{Acknowledgements}

The Submillimeter Array is a joint project between the Smithsonian Astrophysical Observatory and the Academia Sinica Institute of Astronomy and Astrophysics and is funded by the Smithsonian Institution and the Academia Sinica. \swiftbat\/ transient monitor results are provided by the \swiftbat\/ team. This research has made use of \maxi\/ data provided by RIKEN, JAXA and the \maxi\/ team. TSA acknowledges support through the Russian Government Program of Competitive Growth of Kazan Federal University. We thank Sebastian Heinz and Chris Fragile for useful discussions.

\bibliographystyle{aa}

\bibliography{references}

\end{document}